
%
%

\input phyzzx

\overfullrule=0pt
\hoffset=0pt
\hsize=6.5in
\voffset=0in
\vsize=9.0in

\def\L{{\cal L}}
\def\tr{{\rm tr}}
\def\ri{\rightarrow}
\def\mpi{m_{\pi^b}}

\def\M{{\cal M}}
\def\U{{\cal U}}
\def\half{{1\over 2}}
\def\shalf{{1\over\sqrt{2}}}
\def\Q{{\cal Q}}
\def\k{\tilde{k}}
\def\K{{_K}}
\def\ov{\overline}
\def\ep{\varepsilon}
\def\phys{{^{\rm phys}}}
\def\qmu{{^{\rm QM}}}
\def\qmd{{_{\rm QM}}}

\font\el=cmbx10 scaled \magstep2
{\obeylines
\hfill CLNS 93/1189
\hfill IP-ASTP-01-93
\hfill ITP-SB-93-03
\hfill December, 1993}

\vskip 0.5 cm

\centerline {{\el Corrections to Chiral Dynamics of Heavy Hadrons:}}
\centerline {{\el SU(3) Symmetry Breaking}}

\medskip
\bigskip

\centerline{\bf Hai-Yang Cheng$^{a,d}$, Chi-Yee Cheung$^a$, Guey-Lin Lin$^a$,}
\centerline{\bf Y. C. Lin$^b$, Tung-Mow Yan$^{c}$, and Hoi-Lai Yu$^a$}

\medskip
\centerline{$^a$ Institute of Physics, Academia Sinica, Taipei,}
\centerline{Taiwan 11529, Republic of China}

\medskip
\centerline{$^b$ Physics Department, National Central University,
Chung-li,}
\centerline{Taiwan 32054, Republic of China}

\medskip
\centerline{$^c$ Floyd R. Newman Laboratory of Nuclear Studies, Cornell
University}
\centerline{Ithaca, New York 14853, USA }

\medskip
\centerline{$^d$ Institute for Theoretical Physics, State University of
New York}
\centerline{Stony Brook, New York 11794, USA}

\medskip
\bigskip

\centerline{\bf Abstract}

   In previous publications we have analyzed the strong and electromagnetic
decays of heavy mesons and heavy baryons in a formalism which
incorporates heavy-quark and chiral symmetries. There are  two
possible symmetry-breaking effects on the chiral dynamics of heavy hadrons:
the finite-mass effects from  light
quarks and the $1/ m_Q$ corrections from  heavy quarks.
In the present paper, chiral-symmetry-breaking effects are studied and
applications to various strong and radiative decays of heavy hadrons are
illustrated. SU(3) violations induced by chiral loops in the radiative decays
of charmed mesons and charmed baryons are compared with those predicted by the
constituent quark model. In particular, available data for $D^*$ decays favor
values of the parameters in chiral perturbation theory which
give predictions for $D^*$ decays close to the quark model results except for
the $D^{*+}_s$. Implications are discussed.

\endpage
\noindent{\bf 1.~~~Introduction}
\vskip 0.4 cm
   The dominant decay modes of many heavy hadrons, which contain a heavy quark,
are strong decays with one soft pion emission and/or electromagnetic decays.
This is a consequence of the heavy quark symmetry [1,2] of QCD: mass
differences
among the different spin multiplets of the ground-state heavy mesons and
heavy baryons are generally small. An ideal framework for studying the
low-energy dynamics of heavy hadrons is provided by the formalism in which
the heavy quark symmetry and the chiral symmetry are synthesized [3-10].
However, symmetry considerations alone in general do not give any quantitative
predictions unless further assumptions are made. Fortunately, all the unknown
parameters in the Lagrangian  depend only on the
light quarks and are calculable from the nonrelativistic quark model. In
Refs.[3,9,10] (for later convenience, Ref.[3] and Ref.[10] will be denoted as
paper I and paper II respectively, henceforth), we have explored in detail
the predictions of this theoretical formalism on strong decays, radiative
decays, and heavy-flavor-conserving nonleptonic decays.

   In this and a preceding paper, we would like to examine
various symmetry-breaking corrections to the strong and radiative decays
of heavy mesons and baryons.
There are two different kinds of symmetry-breaking effects on the chiral
dynamics of heavy hadrons: the finite-mass effects from the light quarks and
the $1/m_Q$ corrections from the heavy quarks. In paper II, we have already
incorporated
one of the $1/m_Q$ effects, namely the magnetic moment of the heavy quark, into
the formalism for describing the electromagnetic (M1) decay of heavy hadrons.
This is because the charmed quark is not particularly heavy, and hence the
contribution due to its magnetic moment cannot be safely neglected.

   There are two strong motivations, among others, for promoting a
systematic study of both the $1/m_Q$ corrections and the effects of chiral
symmetry breaking. First, we have calculated in paper II the decay rates of
$D^*\ri
D\gamma$. When combined with our prediction for the strong decays $D^*\ri D\pi$
given in paper I, we are able to predict the branching ratios for the $D^*$
decays. Agreement is excellent between theory and the most recent experiment
of CLEO II
[11]. Nevertheless, our predicted total width for $D^{*+}$ is $\Gamma_{\rm
tot}(D^{*+})=150$ keV,
\foot{The difference between this number (see Table I) and the result 141 keV
obtained in Ref.[10] is due to the fact that here we have used a more accurate
pion momentum.}
 which is to be compared with the upper limit $\Gamma
_{\rm tot}(D^{*+})<131$ keV published by the ACCMOR Collaboration [12].
We thus urge the experimentalists to perform more precision measurments of the
$D^*$ total width. Therefore, it becomes urgent to analyze the
symmetry-breaking effects on the strong decays $D^*\to D\pi$ and the radiative
decays $D^*\to D\gamma$. This is particularly so should the aforementioned
upper limit for $\Gamma_{\rm tot}(D^{*+})$ be confirmed by future experiments.
Secondly, to the lowest order chiral Lagrangian, there exist several SU(3)
relations among the radiative decay amplitudes of heavy mesons and heavy
baryons, for example, $A(D^{*+}\to D^+\gamma)=A(D_s^{*+}\to D^+_s\gamma)$.
Therefore, observation of different rates for the decays $D^{*+}\to D^+
\gamma$ and $D_s^{*+}\to D^+_s\gamma$ (after taking into account mass
differences between $D^{+(*)}$ and $D^{+(*)}_s$) will clearly signal the
SU(3) flavor symmetry
breaking due to the light current quark masses. In paper II, the magnetic
moments of heavy hadrons are related to the magnetic moments of the
constituent quarks in the nonrelativistic quark model. In this approach, SU(3)
violations in radiative (M1) decays arise
from the constituent quark mass differences. In the present work, we shall
follow the orthodox approach of chiral perturbation theory to treat
SU(3)-breaking effects: The coupling constants in the Lagrangian are treated
to be invariant and all SU(3) violations of interest are induced by
chiral loops. A comparison of these two different approaches of SU(3) breaking
is given in the end. Existing data for the $D^*$ decays, though limited,
favor a set of parameters such that chiral perturbation theory and the quark
model give similar results for the $D^*$ decays, but not for $D^{*+}_s\ri
D_s^+\gamma$. In addition, even if the upper limit for the total width of
$D^{*+}$ [12] is to be taken seriously, it does not pose a difficulty in
the chiral perturbation theory approach.
For the heavy baryons, lack of any data prevents us from making
a meaningful search for an optimum choice of parameters in chiral
perturbation theory.

Since both symmetry breaking effects require a careful and thorough study,
we focus on the SU(3) symmetry breaking in this paper. A detailed
investigation of $1/m_Q$ corrections has been carried out in a preceding paper
[13]. Schematically, the general effective chiral Lagrangian in chiral
perturbation theory (ChPT) involving heavy hadrons has the chiral expansion
\foot{For pure Goldstone-boson fields, the chiral expansion of chiral
Lagrangians has the familiar form
$$\L=\,\L_2+\L_4+\L_6+\cdots.$$
In this case, higher order chiral terms are suppressed by powers of
$p^2/\Lambda^2_\chi$ or $m^2/\Lambda^2_\chi$.}
$$\L=\,\L_1+\L_2+\L_3+\cdots,\eqno(1.1)$$
where the subscript denotes the sum of the number of derivatives acting on the
Goldstone fields and the velocity-dependent heavy hadron fields,
and the power of Goldstone-boson mass squared in the Lagrangian. (Recall
that the lowest-order chiral-symmetry-breaking terms are linear in light quark
masses and hence quadratic in Goldstone-boson masses.)
The higher-order chiral Lagrangians are suppressed by powers
of $p/\Lambda_\chi$ or $m^2/\Lambda_\chi^2$, where $p$ ($m$) is the
momentum (mass) of the Goldstone bosons, and $\Lambda_\chi$ is a
chiral symmetry breaking scale $\sim$ 1 GeV [14]. Therefore, perturbation
theory makes sense if $p$ and $m$ are not too large compared to
$\Lambda_\chi$. Chiral corrections of interest usually receive two
contributions: one from the chiral loops generated by $\L_1$, and the other
from the higher order tree Lagrangian term $\L_2$. Loop contributions can be
either finite or divergent. The divergent part typically has the form (see
Appendix A)
$${2\over\epsilon}-\gamma_{_{\rm E}}+\ln 4\pi+1+{1\over 2}\ln{\Lambda^2\over
m^2}\eqno(1.2)$$
in the dimensional regularization scheme, where $\epsilon=4-n$, and $\Lambda$
is an arbitrary renormalization scale. In ChPT, all divergences from chiral
loops induced by $\L_1$
\foot{The quadratic divergence, if exists, usually amounts to
renormalizing $\L_1$. In the dimensional regularization scheme we only focus
on those logarithmic divergences.}
will be absorbed into the counterterms which have the same
structure as that of $\L_2$ [15]. Denoting the bare parameters of $\L_2$ by
$f_i$, the renormalized parameters $f_i^r$ are then given by [16]
$$f_i^r(\Lambda)=\,f_i+{\gamma_i\over 32\pi^2}\left({2\over\epsilon}-\gamma_{_{
\rm E}}+\ln 4\pi+1\right)\Lambda^{\epsilon}\eqno(1.3)$$
with $\gamma_i$ being calculable coefficients. We see that although the
lowest-order chiral Lagrangian $\L_1$ is scale independent, renormalized
higher-order effective Lagrangians do depend on the choice of $\Lambda$,
reflecting the non-renormalizability nature of ChPT. Of course, physical
amplitudes should be independent of the renormalization scale; that is, the
$\Lambda$ dependence from chiral-loop is exactly compensated by the $\Lambda$
dependence of local counterterms in $\L_2$.
As the renormalized parameters $f^r_i$ are unknown and must be
determined from experiment, we will thus concentrate in the present paper
the chiral corrections due to meson loops. Furthermore, we will choose
$\Lambda\sim\Lambda_\chi$ to get numerical estimates of chiral-loop effects.

  Chiral loop corrections to some heavy meson processes have been discussed
by other authors [8, 17-20]. In the heavy baryon sector, the only chiral loop
effects studied so far are those in Ref.[17] on the semileptonic decays.
In the present paper, we attempt to make a systematic and complete
investigation of chiral-loop corrections to the strong and
electromagnetic decays of heavy hadrons. They arise from the
chiral-symmetry-breaking terms in (1.1) and hence vanish in chiral limit.
The leading chiral-loop effects we found are nonanalytic in the forms of $m/
\Lambda_\chi$ and $(m^2/\Lambda^2_\chi)\ln(\Lambda^2/ m^2)$ (or $m_q^\half$ and
$m_q\ln m_q$, with $m_q$ being the light quark mass). Furthermore, they amount
to finite-light-quark-mass corrections to the coupling constant, say $g$,
in $\L_1$. Schematically,
$$g_{\rm eff}=\,g\left[1+{\cal O}\left({m\over\Lambda_\chi}\right)+{\cal O}
\left({m^2\over\Lambda_\chi^2}\ln{\Lambda^2\over m^2}\right)+{\cal O}(m^2)
\tilde{g}^r(\Lambda)\right],\eqno(1.4)$$
where $\tilde{g}^r(\Lambda)$ is the relevant renormalized coupling constant in
$\L_2$; that is, the chiral-loop and $\L_2$ contributions have the same
structure as the $g$ term in $\L_1$ except that they vanish in chiral limit.
This point will be elaborated on again in Section 2. In this work we shall
only keep the nonanalytic loop effects.

The present paper is
organized as follows.  In Section 2 we calculate chiral corrections to the
strong decay $P^*\to P\pi$ and the radiative decay $P^*\to P\gamma$, where
$P^*$ and $P$ refer to $1^-$ and $0^-$ ground-state vector and pseudoscalar
mesons, respectively.
As noted earlier, the $1/m_Q$ effect due to the magnetic moment
of the heavy quark is included. Similar chiral-loop calculations are
presented for heavy baryons in Section 3, except there we adopt
velocity-dependent ``superfields'' which combine spin-$\half$ and spin-${
3\over 2}$ sextet baryon fields together [6,17].  Computation becomes much
simplified in this compact notation. A by-product of our investigation
of Sections 2 and 3 is the confirmation at the one loop level of an exact
QCD result (see [13] for example) that the coupling constants due to heavy
quarks in the M1 transitions of both heavy mesons and baryons are not modified
by the light quark dynamics.
In Section 4 we consider
applications of our results to the strong and electromagnetic decays of charmed
mesons and charmed baryons. SU(3) violation induced by chiral loops for the
radiative decays is compared with that predicted by the nonrelativistic quark
model. Section 5 contains conclusions and outlook. Two Appendices are devoted
to some technical details.

\endpage

\noindent{\bf 2.~~~SU(3)-Symmetry-Breaking Corrections to the Chiral
Dynamics of Heavy Mesons}
\vskip 0.4 cm
   In this section we shall study the chiral-symmetry-breaking effects on the
strong
and electromagnetic decays of heavy mesons. To begin with, we recall the
lowest-order gauge invariant chiral Lagrangians (I.2.20) and (II.2.19) for
heavy mesons
$$\eqalign{\L^{(1)}_{v,PP^*}&=-2iM_{P^*}P(v)v\cdot DP^{\dagger}(v)+2iM_{P^*}P^
{*\mu}(v)v\cdot D P_{\mu}^{*\dagger}(v) \cr
&+\Delta M^2P(v)P^{\dagger}(v) +f\sqrt{M_PM_{P^*}}\,[P(v){\cal A}^{\mu}P_{\mu}^
{*\dagger}(v)+P_{\mu}^*(v) {\cal A}^{\mu}P^{\dagger}(v)] \cr
&+2igM_{P^*}\,\epsilon_{\mu\nu\lambda\kappa}
P^{*\mu}(v)v^{\nu}{\cal A}^{\lambda}P^{*\kappa\dagger}(v), \cr}\eqno(2.1)$$
and
$$\eqalign{\L^{(2)}_{v,PP^*}&=\sqrt{M_PM_{P^*}}\epsilon_{\mu\nu
\alpha\beta}v^{\alpha}P^{*\beta}(v)[{1\over 2}d(\xi^{\dagger}
{\cal Q}\xi
+\xi{\cal Q}\xi^{\dagger})+d^{\prime}{\cal Q}^{\prime}]
F^{\mu\nu}P^{\dagger}(v)+h.c.\cr
&+id^{''}M_{P^*}F_{\mu\nu}P^{*\nu}(v)[\gamma {\cal Q}^{\prime}
-{1\over 2}(\xi^{\dagger}{\cal Q}\xi
+\xi{\cal Q}\xi^{\dagger})]P^{*\mu\dagger}(v),\cr}\eqno(2.2)$$
with $\Delta M^2=M^2_{P^*}-M^2_P$,
$$\eqalign{
P_{\mu\nu}^{*\dagger}= &\,D_\mu P_\nu^{*\dagger}-D_\nu P_\mu^{*\dagger}, \cr
P_{\mu\nu}^*= &\,D_\mu P_\nu^*-D_\nu P^*_\mu,  \cr}\eqno(2.3)$$
where $P$ and $P^*$ denote the ground-state $0^{-}$ and $1^{-}$ heavy mesons
respectively which contain a heavy quark $Q$ and a light antiquark $\bar{q}$,
${\cal V}_\mu$ and ${\cal A}_\mu$ are the respective chiral vector and axial
fields (see paper II for more detail), $\Q={\rm diag}(
{2\over 3},-{1\over 3},-{1\over 3})$ is the charge matrix for the light
$u,~d$ and $s$ quarks, $\Q'$ (or $e_Q$) is the charge of the heavy quark,
$\xi=\exp({iM\over\sqrt{2}f_0})$ with the unrenormalized decay constant $f_0$
to be determined later, and $M$ is the meson matrix of Goldstone boson fields
$$M\equiv\sum_a{\lambda^a\pi^a\over\sqrt{2}}=
\left(\matrix{ {\pi^0\over\sqrt{2}}+{\eta\over\sqrt{6}} & \pi^+ & K^+  \cr
\pi^- & -{\pi^0\over\sqrt{2}}+{\eta\over\sqrt{6}} & K^0   \cr
K^- & \ov{K}^0 & -\sqrt{2\over 3}\eta  \cr}\right),\eqno(2.4)$$
where $\lambda$'s are the Gell-Mann matrices normalized by $\tr(\lambda^a
\lambda^b)=2\delta^{ab}$. It should be stressed that the Lagrangians (2.1) and
(2.2) are expressed in terms of velocity-dependent heavy meson fields.

The covariant derivative $D_{\mu}$ in Eqs.(2.1-2.3) is defined by
$$\eqalign{
D_\mu P= &\,\partial_\mu P+{\cal V}_\mu^*P+ieA_\mu(\Q'P-P\Q),  \cr
D_\mu P^\dagger_\nu= &\,\partial_\mu P^\dagger_\nu+{\cal V}_\mu P^\dagger_\nu
-ieA_\mu(P^\dagger_\nu \Q'-\Q P^\dagger_\nu),  \cr}\eqno(2.5)$$
with ${\cal V_{\mu}}={1\over 2}
[\xi^{\dagger}D_{\mu}\xi+\xi(D_{\mu}\xi)^{\dagger}].$
The covariant derivative in ${\cal V}_\mu$ contains the photon field $A_\mu$ to
incorporate the electromagnetic interactions of the Goldstone bosons;
explicitly
$$D_\mu\xi=\,\partial_\mu\xi+ieA_\mu[\Q,~\xi].\eqno(2.6)$$
In Eq.(2.2), $v$ is the velocity of heavy mesons, the two coupling constants
$d\Q$ and $d''\Q$ can be related to the
magnetic moments of light constituent quarks in the quark model.
The universal coupling constants $d$ and $d''$ are independent of the heavy
quark masses and species. We have also included the $d'\Q'$ and $\gamma\Q'$
terms to account for the corrections due to the heavy quark masses when
$m_Q\not= \infty$.
   The coupling constants $f$ and $g$ in (2.1), $d$ and $d''$ in (2.2)
are related by heavy quark symmetry as [3,10]
$$f=\,2g,~~~d''=-2d.\eqno(2.7)$$
Moreover, the couplings $d'$ and $d''\gamma$ are fixed by heavy quark symmetry
to be [10]
$$d'=-{e\over 2m_Q},~~~d''\gamma=\,{e\over m_Q}.\eqno(2.8)$$

We shall see that from the lowest-order Lagrangians given by (2.1) and (2.2),
there is an SU(3) prediction: $A(D^{*+}\ri D^+\gamma)=A(D^{*+}
_s\ri D^+_s\gamma)$. Therefore, observation of different rates for the
radiative decays of $D^{*+}$ and $D^{*+}_s$ (after taking into account the
 mass difference of $D^{+(*)}$ and $D^{+(*)}_s$) will evidently signal the
$SU(3)$-breaking effects induced by the light quark masses. In chiral
perturbation theory the masses of light quarks are treated as a small
perturbation and SU(3) violation in radiative decays is induced by
chiral loops. In paper II, the
unknown coupling constants $d$ and $d''$ in (2.2)  are derived from the
non-relativistic quark model and they are related to the masses of the
constituent $u,~d$ and $s$ quarks. As a consequence, SU(3) violation is
already incorporated into the ``effective'' couplings $d$ and
$d''$ in paper II. In the present work, we shall adopt the orthodox approach
of chiral perturbation theory for treating SU(3)-breaking effects. As will be
shown, leading chiral corrections have nonanalytic dependence on $m_q$ of
the form $m_q^\half$ or $m_q\ln m_q$.
In Sec. 4 the above two different approaches for SU(3) violation in radiative
decays of heavy hadrons will be compared.

The chiral symmetry breaking terms are given by
$$\eqalign{\L_{\rm CSB}=&{f_0^2\over 2}\,B_0\,\tr(\M^\dagger\Sigma +\Sigma^{
\dagger}\M)\cr   +&\alpha_1M_PP(v)(\xi \M^\dagger \xi +\xi^{\dagger} \M \xi^{
\dagger})P^{\dagger}(v)  +\alpha_2M_{P^*}P^*_{\mu}(v)(\xi \M^\dagger \xi +\xi^
{\dagger}\M \xi^{\dagger})P^{*\mu}(v)\cr
+&\alpha_3M_PP(v)P^{\dagger}(v)\tr( \M^\dagger \Sigma +\Sigma^{\dagger} \M)
+\alpha_4M_{P^*}P^{*\mu}(v)P^{*\dagger}_{\mu}(v)
\tr( \M^\dagger\Sigma +\Sigma^{\dagger} \M),\cr}\eqno(2.9)$$
where $\Sigma=\xi^2$, $B_0=-4\VEV{\bar{q}q}/f_0^2$ characterizes the
spontantous breaking of chiral symmetry, and $\M$ is a light quark mass matrix
$$\M=\pmatrix{m_u&0&0\cr 0&m_d&0\cr 0&0&m_s\cr}.\eqno(2.10)$$
The unknown dimensionless parameters $\alpha_i$ are expected to be of order
unity (see Sec. 4 for more discussion).
Each symmetry breaking term in (2.9) transforms either as $(\bar 3, 3)$
or $(3, \bar 3)$ under $SU(3)_L\times SU(3)_R$. As the symmetry breaking
effect is purely induced by the light quark masses, $\L_{\rm CSB}$ should be
independent of the heavy quark mass and spin. By comparing (2.9) with
(2.1), this immediately gives $\alpha_1=-\alpha_2$ and
$\alpha_3=-\alpha_4$. As one can see from (2.9), $\L_{\rm CSB}$ does not
break the heavy-quark-symmetry (HQS) relations given by Eq.(2.7).
 In fact, chiral symmetry breaking should preserve HQS as light meson
interactions have nothing to do with the heavy quarks.

    We shall first calculate the chiral-loop contributions to the
strong decay $P^*_i\ri P_j+\pi^a$ as depicted in Fig. 1, where $P_i$ denotes
$(Q\bar{u},~Q\bar{d},~Q\bar{s})$ and likewise for $P^*_i$.
The tree amplitude can be read out from (2.1) to be
$$A[P^*_i\ri P_j\pi^a(q)]_{\rm tree}=\,{g\over f_0}\sqrt{M_PM_{P^*}}
(\lambda^a)_{ij}(\ep^*\cdot q).\eqno(2.11)$$
As discussed in the Introduction, the loop-induced logarithmic divergence is
absorbed into the chiral-symmetry-breaking counterterms
$$aP(v)(\xi{\cal M}^\dagger\xi+\xi^\dagger{\cal M}\xi^\dagger){\cal A}^\mu
P^{*\dagger}_\mu(v)+bP(v){\cal A}^\mu(\xi{\cal M}^\dagger\xi+\xi^\dagger{\cal
M}\xi^\dagger)P^{*\dagger}_\mu(v)+h.c.\eqno(2.12)$$
Obviously, in the above expression, we should take $\xi=1$ in $(\xi M^\dagger
\xi+\xi^\dagger M\xi^\dagger)$ in order to describe the decay $P^*\ri
P\pi$. As a consequence, the higher order contribution due to (2.12)
has the same structure as that of (2.11) except that the former vanishes in
chiral limit. As stressed in the Introduction, we will not consider
higher-order Lagrangian effects in the present paper because of the unknown
parameters $a$ and $b$ in (2.12). The sum of all chiral loop
contributions shown in Fig. 1 gives rise to the effective coupling constant
$$g_{{\rm eff}}=\,g{\sqrt{Z_2(P)Z_2(P^*)Z_2(\pi^a)}\over Z_1}=\,g{Z_2(P)
\sqrt{Z_2(\pi^a)}\over Z_1},\eqno(2.13)$$
where $Z_1$ and $Z_2$ are the vertex and wave-function renormalization
constants, respectively,
and the HQS relation $Z_2(P^*)=Z_2(P)$ has been applied.  For simplicity,
wave-function renormalization and mass counterterms of the heavy meson and
Goldstone boson are not explicitly shown
in Fig. 1, but the usual mass renormalization procedure is to be understood.
   To evaluate the $Z_2$ renormalization constant, we note that
the self-energy amplitude of, for example $P_i$, has the expression
$$-i\Pi(\k)=-{g^2\over 2f_0^2}M_P\sum_b(\lambda^b\lambda^b)_{ii}\int {d^4l
\over
(2\pi)^4}\,{(g_{\mu\nu}-v_\mu v_\nu)l^\mu l^\nu\over (l^2-\mpi^2+i\epsilon)[v
\cdot(l+\k)+i\epsilon]},  \eqno(2.14)$$
where $\k$ is the residual momentum defined by $P_\mu=m_Qv_\mu+\k_\mu$, and
$\mpi$ is the mass of the Goldstone boson $\pi^b$ in the loop. (For
notation consistency, we will follow papers I and II to denote the momenta of
the Goldstone boson and photon by $q$ and $k$, respectively.)
With the help of Eq.(A4) and the relation
$$\Pi(\k)=\delta m^2-2(Z_2^{-1}-1)(v\cdot \k)M_P,\eqno(2.15)$$
we obtain
$$\delta m^2(P_i)=-{g^2\over 16\pi}\sum_b(\lambda^b\lambda^b)_{ii}{\mpi^3\over
f_{0}^2}M_{P_i},\eqno(2.16)$$
and
$$Z_2(P_i)=\,1+{3g^2\over 64\pi^2}\sum_b(\lambda^b\lambda^b)_{ii}{\mpi^2\over
f_{0}^2}\ln{\Lambda^2\over\mpi^2},\eqno(2.17)$$
where $\Lambda$ is an arbitrary renormalization scale. As discussed in the
Introduction, divergences from chiral loops are absorbed into the
unrenormalized parameters of higher-derivative chiral Lagrangians which are not
written down here.
It is straightforward to check that the HQS relation $Z_2(P^*_i)=Z_2(P_i)$
holds. Note that the SU(3) invariant masses $M_P$ and $M_{P^*}$ in Eq.(2.11)
are unrenormalized masses; they are connected to the physical masses through
the relations:
$$\eqalign{ M^2_{_{\rm phys}}(P_i) =&\,M^2_{P}-2\alpha_1M_P{\cal M}_{ii}-2
\alpha_3M_P\tr{\cal M}+\delta m^2(P_i),   \cr
M^2_{_{\rm phys}}(P_i^*) =&\,M^2_{P^*}-2\alpha_1M_{P^*}{\cal M}_{ii}-2\alpha_3
M_{P^*}\tr{\cal M}+\delta m^2(P_i^*).   \cr}\eqno(2.18)$$
It follows from Eqs. (2.9), (2.16) and (2.18) that
the mass splitting of, say $P_3=(Q\bar{s})$ and $P_1=(Q\bar{u})$, is given by
$$M_{P_3}^\phys-M_{P_1}^\phys=\,-\alpha_1(m_s-m_u)-{g^2\over 32\pi}\sum_b[(
\lambda^b\lambda^b)_{33}-(\lambda^b\lambda^b)_{11}]{\mpi^3\over f_{0}^2}.
\eqno(2.19a)$$
Likewise, for heavy vector mesons
$$M_{P^*_3}^\phys-M_{P^*_1}^\phys=\,M_{P_3}^\phys-M_{P_1}^\phys.
\eqno(2.19b)$$
Consequently, a meaurement of the
heavy-meson mass differences will provide information on the parameter
$\alpha_1$. We will come back to this in Sec. 4.

   From the strong-interaction chiral Lagrangian for the Goldstone bosons
$$\L=\,{f^2_0\over 4}\tr\left(\partial_\mu\Sigma^\dagger\partial^\mu\Sigma
\right)+{f_0^2\over 4}\tr\left(\ov{\cal M}^\dagger\Sigma+\ov{\cal M}\Sigma
^\dagger\right)\eqno(2.20)$$
with $\ov{\cal M}={\rm diag}(m^2_\pi,~m^2
_\pi,~2m^2_\K-m^2_\pi)$, the wave-function renormalization constants for the
light pseudoscalar mesons are found to be
$$\eqalign{ \sqrt{Z_2(\pi)} =& 1-{1\over 3}(2\epsilon_\pi+\epsilon_\K),  \cr
\sqrt{Z_2(K)} =& 1-{1\over 3}({3\over 4}\epsilon_\pi+{3\over 2}\epsilon_\K
+{3\over 4}\epsilon_\eta),   \cr
\sqrt{Z_2(\eta)} =& 1-{1\over 3}(3\epsilon_\K),  \cr}\eqno(2.21)$$
with
$$\eqalign{ \epsilon_\pi= &\,{1\over 32\pi^2}\,{m_\pi^2\over f_0^2}\ln{
\Lambda^2\over m^2_\pi},   \cr
\epsilon_{_K}= &\,{1\over 32\pi^2}\,{m_\K^2\over f_0^2}\ln{
\Lambda^2\over m^2_\K},   \cr
\epsilon_\eta= &\,{1\over 32\pi^2}\,{m_\eta^2\over f_0^2}\ln{
\Lambda^2\over m^2_\eta}.   \cr}\eqno(2.22)$$

   As to the Feynman diagrams Figs. 1a-1e, the amplitude of the
vertex diagrams Figs. 1a and 1b reads
$$A(1a+1b)={g^3\over 64\pi^2}\sqrt{M_PM_{P^*}}\sum_b(\lambda^b\lambda^a
\lambda^b)_{ij}{\mpi^2\over f_0^3}\ln{\Lambda^2\over
\mpi^2}(\ep^*\cdot q),\eqno(2.23)$$
where we have applied Eq.(A3). To evaluate the seagull graphs of Figs. 1c-1d,
we note that the Feynman rules for the vertices $PP\pi^a\pi^b$ and $P^*P^*
\pi^a\pi^b$ are
obtained from the kinematic terms in Eq.(2.1) together with Eqs.(2.3-2.6) and
$${\cal V}_\mu=\,{1\over 4f_0}(M\partial_\mu M-\partial_\mu MM)+\cdots,
\eqno(2.24)$$
where $M$ is the meson matrix given by Eq.(2.4). The relevant Feynman rules
are
$$-{i\over 4f_0^2}\,M_P(\lambda^a\lambda^b-\lambda^b\lambda^a)v\cdot(q_a
-q_b)\eqno(2.25)$$
for the vertex $PP\pi^a\pi^b$, and
$$ {i\over 4f_0^2}\,M_P(\lambda^a\lambda^b-\lambda^b\lambda^a)
v\cdot(q_a-q_b)(\ep^*\cdot\ep'^*)  \eqno(2.26)$$
for $P^*P^*\pi^a\pi^b$. Therefore,
both vertices are proportional to $v\cdot(q_a
-q_b)$. As a consequence, seagull diagrams Figs. 1c and 1d vanish since the
required $\ep^*\cdot q$ expression cannot be generated from the seagull
amplitudes in which the linear $q$ dependence is always in the $v\cdot q$ form.
Beyond the heavy quark limit, seagull graphs do contribute but they are
suppressed by factors of $1/M_{P^*}$ relative to the tree amplitude.
For example, to order $1/M_{P^*}$, the following $P^*P^*\pi^a\pi^b$ vertex
$$-{i\over 8f_0^2}(\lambda^a\lambda^b-\lambda^b\lambda^a)M_P[(q_a-q_b)\cdot
\ep'^*(v\cdot \ep^*)-(q_a-q_b)\cdot\ep^*(v\cdot\ep'^*)]$$
are no longer vanishing as $v\cdot\ep^*\neq 0$ and $v\cdot\ep'^*\neq 0$.

   The 5-point vertex in Fig. 1e is generated from the Lagrangian (2.1) by
expanding ${\cal A}_\mu$ to the third power of $M$
$${\cal A}_\mu=\,-{1\over \sqrt{2}f_0}\,\partial_\mu M+
{1\over 12\sqrt{2}f_0^3}\left[(\partial_\mu M)M^2-2M(\partial
_\mu M)M+M^2\partial_\mu M\right]+\cdots.\eqno(2.27)$$
A straightforward calculation yields
$$A(1e)=\,{g\over f_0}\sqrt{M_PM_{P^*}}(\lambda^a)_{ij}(\ep^*\cdot q)
\times
\cases{ {1\over 3}(2\epsilon_\pi+\epsilon_\K), & {\rm for}~$P_i^*\ri P_j+\pi$,
 \cr {1\over 3}({3\over 4}\epsilon_\pi+{3\over 2}\epsilon_\K
+{3\over 4}\epsilon_\eta), & {\rm for}~$P_i^*\ri P_j+K$,  \cr {1\over 3}
(3\epsilon_\K), & {\rm for}~$P_i^*\ri P_j+\eta$.  \cr}\eqno(2.28)$$
Since the vertex corrections are lumped into the renormalization constant
$Z_1$ by
$$A(1a+1b+1c+1d+1e)=\,(Z_1^{-1}-1)A(P^*_i\ri P_j\pi^a)_{\rm tree},\eqno(2.29)$$
it is clear from Eqs.(2.13), (2.21) and (2.28) that the contribution due to
the 5-point vertex diagram Fig. 1e is exactly compensated by the wave-function
renormalization of the Goldstone bosons. As a consequence, we have effectively
$$g_{\rm eff}=\,g{Z_2(P)\over Z_1(1a+1b)},\eqno(2.30)$$
with
$$ Z_1(1a+1b)=\,1 -{g^2\over 2}\sum_b{(\lambda^b\lambda^a
\lambda^b)_{ij}\over(\lambda^a)_{ij}}\,\epsilon_{\pi^b}.\eqno(2.31)$$
Using the relations
$$\eqalign{ \sum_b(\lambda^b\lambda^b)_{11}f(\pi^b) =& \sum_b(\lambda^b\lambda
^b)_{22}f(\pi^b)=\,3f(\pi)+2f(K)+{1\over 3}f(\eta),   \cr
\sum_b(\lambda^b\lambda^b)_{33}f(\pi^b) =&\, 4f(K)+{4\over 3}f(\eta),  \cr}
\eqno(2.32)$$
it follows from Eq.(2.17) that the individual wave function renormalization
constants for heavy mesons are given by
$$\eqalign{ &Z_2(P_1)=\,Z_2(P_2)=\,1+g^2\left({9\over 2}
\epsilon_\pi+3\epsilon_{_K}+{1\over 2}\epsilon_\eta\right),  \cr
& Z_2(P_3)=\,1+2g^2(3\epsilon_{_K}+\epsilon_\eta).  \cr}\eqno(2.33)$$
In Sec. 4 the above results will be applied
to the strong decay $D^*\ri D+\pi$. There we shall see that $Z_2$ plays an
essential role and the pion contribution is not negligible in spite of its
small mass.

   Thus far we have expressed all the results in terms of the unrenormalized
decay constant $f_0$. It can be related to the physical decay constants
through the relations [16]:
$$\eqalign{ f_\pi =&\,f_0\left(1+2\epsilon_\pi+\epsilon_\K+4\,{2m^2_\K+m_\pi^2
\over f_0^2}L_4^r+4{m_\pi^2\over f_0^2}L_5^r\right),  \cr
f_\K =&\,f_0\left(1+{3\over 4}\epsilon_\pi+{3\over 2}\epsilon_\K+{3\over 4}
\epsilon_\eta+4\,{2m^2_\K+m_\pi^2\over f_0^2}L_4^r+2\,{m_\pi^2\over f_0^2}{m_s+
\hat{m}\over \hat{m}}L_5^r\right),  \cr
f_\eta =&\,f_0\left(1+3\epsilon_\K+4\,{2m^2_\K+m_\pi^2
\over f_0^2}L_4^r+{4\over 3}{m_\pi^2\over f_0^2}{2m_s+\hat{m}\over \hat{m}}
L_5^r\right),   \cr}\eqno(2.34)$$
where $\hat{m}=(m_u+m_d)/2$, and the counterterm contributions denoted by the
renormalized coupling constants $L_4^r$ and $L_5^r$ (see Ref.[16] for
notation) are included in Eq.(2.34). Note that the one-loop logarithmic
corrections to the decay constants, the pion wave-function renormalization and
the 5-point vertex diagram all share the
same structure as they come from the same chiral loop diagram. The
couplings $L_4^r$ and $L_5^r$ of the four-derivative chiral Lagrangian
are dependent of the renormalization scale $\Lambda$. However, physical decay
constants can be verified to be independent of the choice of $\Lambda$, as
they should be [16]. Using the empirical values
$$L_4^r(\Lambda=m_\eta)=(0\pm 0.5)\times 10^{-3},~~~L_5^r(\Lambda=m_\eta)=(
2.2\pm 0.5)\times 10^{-3}\eqno(2.35)$$
obtained in Ref.[16] and the experimental value $f_\pi=93$ MeV [21], we find
$$f_0\simeq\,86\,{\rm MeV}.\eqno(2.36)$$

   We make a digression here to comment on the $1/m_Q$ effects.
In the heavy quark effective theory, $1/m_Q$ corrections can be systematically
studied by including the following higher-dimensional operators
$$\eqalign{ O_1=&\,{1\over 2m_Q}\bar{h}_v^{(Q)}(iD)^2h_v^{(Q)}, \cr
 O_2=&\,{1\over 2m_Q}\bar{h}_v^{(Q)}(-{1\over 2}g_s\sigma_{\mu\nu}G^{\mu\nu})
h_v^{(Q)},  \cr}\eqno(2.37)$$
where $h_v^{(Q)}$ is a velocity-dependent heavy quark field.
For the case of $P^*\to P\pi$ decays, there are two different kinds of $1/m_Q$
corrections. One arises from the consideration of the subleading decay
amplitude induced by the operators $O_1$ and $O_2$
$$A_{s.l.}[P^*(v,\varepsilon^*)\rightarrow P(v')\pi^a(q)]
=\,{1\over f_0}q^{\mu}\bra{P(v')} iT\int d^4x[O_1(x)+O_2(x)]{\cal A}_{\mu}^
a(0)\ket{P^*(v,\varepsilon^*)}.\eqno(2.38)$$
As shown in the preceding paper [13], the coupling constant $g$ receives
a correction of order $\Lambda_{_{\rm QCD}}/m_Q$; that is,
$$g_{\rm eff}=\,g\left[1+{\cal O}\left({\Lambda_{_{\rm QCD}}\over m_Q}\right)
\right].\eqno(2.39)$$
This $1/m_Q$ effect in chiral limit is elaborated on in more details in
Ref.[13]. The other kind of $1/m_Q$ corrections comes from the combination of
chiral-symmetry and
heavy-flavor-symmetry breakings, e.g., the seagull diagrams. Schematically,
it is expected to be order of
$${1\over (4\pi)^2}{m^2_{\pi^b}\over f^2_{0}}{m_{\pi^b}\over m_Q},
\eqno(2.40)$$
where the factor $1/(4\pi)^2$ is associated with the loop-momentum integration,
and $m_{\pi^b}$ is the mass of the loop meson $\pi^b$.
As this SU(3)-violating $1/m_Q$ effect is very small, we will not pursue
it further.

    We next turn to the electromagnetic decay of heavy mesons. One-loop
contributions to $P^*\ri P+\gamma$ are shown in Fig. 2. The Feynman diagrams
with a photon coupled to the external $P^*$ or $P$ line via charge coupling are
vanishing, as expected since the reaction under consideration involves a
magnetic M1 transition. Therefore, they are not
displayed in Fig. 2. Although chiral-loop effects in the radiative decays
have been recently discussed
by Amundson {\it et al}. [8], only Fig. 2e was analyzed by them.
Before proceeding to compute loop corrections, we note that the
tree amplitude obtained from (2.2) is
$$A[P^*_i(v,\ep^*)\ri P_i+\gamma(k,\ep)]=\rho_i\epsilon_{\mu\nu
\alpha\beta}k^\mu \ep^\nu v^\alpha\ep^{*\beta},\eqno(2.41)$$
with
$$\rho_i=-2\sqrt{M_PM_{P^*}}(e_{i} d+e_Qd')=\,e\sqrt{M_PM_{P^*}}(e_{i}
\beta+e_Q\beta'),\eqno(2.42)$$
and $e_i=Q_{ii}$.
Though the unknown parameter $\beta\equiv -2d/e$
is not fixed by heavy quark symmetry, the other parameter $\beta'\equiv-2d'/e$
is determined in the heavy quark effective theory by the dimension-five
operator
$$ O_3=\,{1\over 2m_Q}\bar{h}_v^{(Q)}(-{1\over 2}ee_Q\sigma_{\mu\nu}F^{\mu\nu})
h_v^{(Q)},\eqno(2.43)$$
with the result [7,8,10]
$$\beta'=\,{1\over m_Q}.\eqno(2.44)$$
It should be stressed that it is important to include this $1/m_Q$ effect due
to the magnetic moment of the heavy quark. As we have seen in paper II, the
charmed quark contribution is significant and largely destructive in the
radiative decays of $D^{*+}$ and $D^{*+}_s$. In the non-relativistic
quark model $\beta$ is related to the constituent quark mass, namely
$\beta=1/M_q$ [10]. Since in the spirit of chiral
perturbation theory SU(3) violation is induced by chiral-symmetry breaking
terms, it is interesting to compare the quark model results
with those of the chiral Lagrangian approach.
To the lowest order, it is evident that $\rho_2=\rho_3$ or
$$A(P^*_2\ri P_2\gamma)=A(P^*_3\ri P_3\gamma).\eqno(2.45)$$
As will be seen, this SU(3) relation is badly broken by chiral-loop effects.

   Referring to Fig. 2 the effective couplings become
$$\eqalign{  (\beta_i)_{\rm eff}=&\,\beta{Z_2(P_i)\over (Z_1)_i}+\delta\beta_i
(2a+2b+2e), \cr  (\beta'_i)_{\rm eff}=&\,\beta'{Z_2(P_i)\over (Z_1')_i},
  \cr}\eqno(2.46)$$
where the subscript $i$ refers to the process $P_i^*\ri P_i\gamma$
 and $Z_1$ accounts for the vertex renormalization due to Figs. 2c-2d and 2f.
We begin with Fig. 2e.
In order to evaluate its amplitude, all the charged meson loops are
explicitly shown in Fig. 3. By virtue of Eq.(A2), we find
\foot{In Ref.[8], $f_0$ is replaced by the physical decay constant $f_\pi$ for
the pion loop and $f_\K$ for the kaon loop. We will not make such a replacement
in our calculation.}
$$\eqalign{\delta\beta_1(2e)= &\,{1\over e_1}\left(-{g^2\over
8\pi}{m_\pi\over f^2_0}-{g^2\over 8\pi}{m_\K\over f^2_0}\right),  \cr
 \delta\beta_2(2e)= &\,{1\over e_2}\left({g^2\over
8\pi}{m_\pi\over f^2_0}\right),  \cr
 \delta\beta_3(2e)= &\,{1\over e_3}\left({g^2\over
8\pi}{m_\K\over f^2_0}\right),  \cr}\eqno(2.47)$$
which are in agreement with Ref.[8].
As for the seagull diagrams Figs. 2a and 2b, we note that the
electromagnetic four-point interactions of heavy mesons are described by
$$\eqalign{
&{-\sqrt{2}eg\over f_0}\epsilon^{\mu\nu\lambda\kappa}A_\mu v_\nu
P^*_\lambda(v)[\Q,~M]P^{*\dagger}_\kappa(v)-i\sqrt{M_PM_{P^*}}\,{\sqrt{2}
eg\over f_0}\,A^\mu  \cr  &\times\Big(P(v)[\Q,~M]P_\mu
^{*\dagger}(v)+P_\mu^*(v)[\Q,~M]P^\dagger(v)\Big),  \cr}\eqno(2.48)$$
where use has been made of
$$\eqalign{ {\cal A}_\mu= &\,{\cal A}_\mu^{(0)}-\half eA_\mu(\xi^\dagger\Q\xi
-\xi\Q\xi^\dagger)   \cr  =&\,-{1\over \sqrt{2}f_0}\partial_\mu M-{ie\over
\sqrt{2}f_0}A_\mu [\Q,~M]+\cdots   \cr}\eqno(2.49)$$
in (2.1) and (2.2). It is clear that the $P^*P\pi^a\gamma$ vertex
is independent of the photon's momentum $k$. As a consequence, Fig. 2b with a
photon emitted from the vertex on the r.h.s.
does not contribute since no linear $k$-dependent terms can be
constructed from the numerator or the denominator after the loop mementa
of the $\pi^b$ and $P^*$ are assigned to be $l$ and $p+l$, respectively.
Since the vertex $P^*P^*\pi^a\gamma$ is proportional to $\epsilon^{\mu\nu
\lambda\kappa}v_\mu \ep_\nu\ep^*_\lambda{\ep'}^*_\kappa$ in the
heavy quark limit, it is easily seen that the other seagull diagram Fig. 2a
also vanishes. Its subleading contribution is of order $(g^2/f_0^2)(1/M_
{P^*})$ and hence negligible.

   We next focus on vertex corrections. With the magnetic $P^*P\gamma$ and
$P^*P^*\gamma$ vertices determined from (2.2), Figs. 2c and 2d lead to
$$Z_1(P^*_i\ri P_i\gamma;2c+2d)=\,1-{g^2\over 2}\sum_b{(\lambda^b\Q\lambda^b)
_{ii}\over \Q_{ii}}\,\epsilon_{\pi^b},\eqno(2.50)$$
and
$$Z'_1(P^*_i\ri P_i\gamma;2c+2d)= \,1+{3\over 2}g^2\sum_b(\lambda^b\lambda^
b)_{ii}\epsilon_{\pi^b}=\,Z_2(P_i),\eqno(2.51)$$
where we have applied (2.17) and the relations (2.7) and (2.8).
The contribution of Fig. 2c due to charge $P^*P^*\gamma$ coupling is of order
$(g^2/32\pi^2)(m_{\pi^b}^2/f_0^2)(m_{\pi^b}/M_P)$ and hence negligible.
Note that because $Z'_1(P^*_i\ri P_i\gamma)$ is exactly compensated by $Z_2(
P_i)$, the parameter $\beta'$ of (2.46)
does not get renormalized; this is a realization
in chiral perturbation theory of the exact QCD result discussed in Ref.[13].

   It remains to compute the amplitude of Fig. 2f. The relevant electromagnetic
5-point vertex obtained from (2.2) is
$$ -{d\over 2f_0^2}\sqrt{M_PM_{P^*}}\,\epsilon_{\mu\nu\alpha\beta}
k^\mu\ep^\nu v^\alpha\ep^{*\beta}[M,~[\Q,~M]].\eqno(2.52)$$
It is easily seen that only charged meson loops contribute; the result is
$$Z_1(P^*_i\ri P_i\gamma;2f)=\,1+{1\over 4}\sum_b[\lambda^b,~[\Q,~\lambda^b]]
_{ii}{\epsilon_{\pi^b}\over \Q_{ii}}.\eqno(2.53)$$
Working out (2.50) and (2.53) explicitly, we find the total vertex corrections
to be:
$$\eqalign{ Z_1(P_1^*\ri P_1\gamma;2c+2d+2f) =&\,1+{1\over e_1}\left[-\epsilon
_\pi-\epsilon_\K+{g^2\over 3}\left(\epsilon_\K-{1\over 3}\epsilon_\eta\right)
\right], \cr    Z_1(P_2^*\ri P_2\gamma;2c+2d+2f) =&\,1+{1\over e_2}\left[
\epsilon_\pi+{g^2\over 3}\left(-{3\over 2}\epsilon_\pi+\epsilon_\K+{1\over 6}
\epsilon_\eta\right)\right],   \cr    Z_1(P_3^*\ri P_3\gamma;2c+2d+2f) =&1+{1
\over e_3}\left[\epsilon_\K+{g^2\over 3}\left(-\epsilon_\K+{2\over 3}\epsilon_
\eta\right)\right].  \cr}\eqno(2.54)$$
Putting everything together, we obtain the following effective couplings for
$P^*_i\ri P_i\gamma$:
$$\eqalign{  \rho_1= &\,e\sqrt{M_PM_{P^*}}\left[{2\over 3}\beta\,
{Z_2(P_1)\over Z_1(P^*_1\ri P_1\gamma)}+{e_Q\over m_Q}-{g^2\over 8\pi}\,{m_
\pi\over f_0^2}-{g^2\over 8\pi}\,{m_\K\over f_0^2}\right],   \cr
 \rho_2= &\,e\sqrt{M_PM_{P^*}}\left[-{1\over 3}\beta\,
{Z_2(P_2)\over Z_1(P^*_2\ri P_2\gamma)}+{e_Q\over m_Q}+{g^2\over 8\pi}
\,{m_\pi\over f_0^2}\right],   \cr
 \rho_3= &\,e\sqrt{M_{P}M_{P^*}}\left[-{1\over 3}\beta\,{Z_2(P_3)\over Z_1(
P_3^*\ri P_3\gamma)}+{e_Q\over m_Q}+{g^2\over 8\pi}\,{m_{_K}
\over f_{0}^2}\right].   \cr   }\eqno(2.55)$$
In the SU(3) limit, the light quark contributions to $\rho$ are in the ratio
$2:-1:-1$ for the electromagnetic decays of $P^*_1,~P^*_2,$ and
$P^*_3$. Evidently, this relation is violated by the wave function and vertex
renormalizations.

\vskip 0.5 cm
\noindent{\bf 3.~~SU(3)-Symmetry-Breaking Corrections to the Chiral
Dynamics of Heavy Baryons}
\vskip 0.4 cm

     In this section we discuss the corrections to the strong and radiative
decays of heavy baryons containing a heavy quark $Q$ and two light quarks.
The two light quarks form either a symmetric sextet {\bf 6} or an
antisymmetric antitriplet {\bf \={3}} in flavor SU(3) space.  We will
denote these spin $\half$ baryons as $B_6$ and $B_{\bar{3}}$
respectively, and the spin ${3\over 2}$ baryon by $B^\ast_6$.  Explicitly,
the baryon matrices read as in paper I
$$ B_{\bar{3}}=\left(\matrix{ 0& \Lambda_Q & \Xi_Q^\half  \cr
-\Lambda_Q & 0 & \Xi_Q^{-\half} \cr  -\Xi_Q^\half & -\Xi_Q^{-\half} & 0 \cr}
\right),\eqno(3.1)$$
$$ B_6 =\left(\matrix{\Sigma_Q^{+1} & \shalf\Sigma_Q^0 & \shalf{\Xi'}_Q^\half
  \cr
\shalf\Sigma_Q^0 & \Sigma_Q^{-1} & \shalf{\Xi'}_Q^{-\half}  \cr  \shalf{\Xi'}_Q
^\half & \shalf{\Xi'}_Q^{-\half} & \Omega_Q  \cr}\right),\eqno(3.2)$$
and a matrix for $B_6^*$ similar to $B_6$. The superscript in (3.1) and
(3.2) refers to the value of the isospin quantum number $I_3$.

To perform chiral-loop calculations, we find that for sextet heavy baryons it
is very convenient to work with the ``superfields'' which combine
spin-${1\over 2}$ and spin-${3\over 2}$ sextet baryon fields into a single
field [6,17]
$$\eqalign{ S^\mu= &\,B_6^{*\mu}-{1\over\sqrt{3}}(\gamma^\mu+v^\mu)\gamma_5
B_6,   \cr  \bar{S}^\mu= &\,\bar{B}_6^{*\mu}+{1\over\sqrt{3}}\bar{B}_6\gamma_5
(\gamma^\mu+v^\mu),  \cr}\eqno(3.3)$$
where $B^*_{6\mu}$ is the Rarita-Schwinger vector-spinor field as appropriate
for spin ${3\over 2}$ particles.
Feynman rules in terms of superfields become much simpler and there are fewer
Feynman diagrams to evaluate.
The most general gauge-invariant chiral Lagrangian for heavy baryons given by
Eq.(II.3.8) reads
$$\eqalign{ \L_B^{(1)}= &\,-i\tr(\bar{S}^\mu v\cdot DS_\mu)+{i\over 2}\tr(\ov
{T}v\cdot DT)+\Delta m\,\tr(\bar{S}^\mu S_\mu)  \cr
&+i{3\over 2}\,g_1\epsilon_{\mu\alpha\beta\nu}\tr(\bar{S}^\mu v^\alpha{\cal A}
^\beta S^\nu)-\sqrt{3}\,g_2\tr(\bar{S}^\mu{\cal A}_\mu T)+h.c.\cr}\eqno(3.4)
$$
in superfield notation, where $T\equiv B_{\bar{3}}$, $\Delta m$ is the mass
splitting between the sextet and antitriplet baryon multiplets (for
simplicity, we will drop the unknown $\Delta m$ term in ensuing loop
calculations), and
$$D_\mu B=\,\partial_\mu B+{\cal V}_\mu B+B{\cal V}_\mu^T+ie\Q'A_\mu B
+ieA_\mu\{\Q,~B\}  \eqno(3.5)$$
for $B=S_\nu,~T$. We wish to stress that the Lagrangian (3.4) is expressed in
terms of velocity dependent baryon fields.
Because of heavy quark symmetry, there are only two
independent coupling constants in the low-energy interactions between the
Goldstone bosons and the heavy baryons. Moreover, the nonrelativistic
quark model predicts that [3]:
$$g_1=\,{1\over 3}g,~~~g_2=-\sqrt{2\over 3}\,g\eqno(3.6)$$
with $g$ being the axial vector coupling constant of a constituent
quark, which is also the coupling constant appearing in the meson Lagrangian
(2.1). Feynman rules can
be easily derived from (3.4). Especially, the $S$ and $T$ propagators
are simply given by $i(-g_{\mu\nu}+v_\mu v_\nu)/(v\cdot \k-\Delta m)$ and
$i/v\cdot \k$,
respectively, where $\k$ is the residual momentum of the heavy hadron.
The lowest order chiral-symmetry-breaking terms now have the form
$$\eqalign{\L_{\rm CSB} = &\,\lambda_1\,\tr[\bar{S}^\mu(\xi\M^\dagger\xi+\xi^
\dagger\M\xi^\dagger)S_\mu]+\lambda_2\,\tr(\bar{S}^\mu S_\mu)\tr(\M^\dagger
\Sigma+\Sigma^\dagger\M)  \cr +&\lambda_3\, \tr[\ov{T}(\xi\M^
\dagger\xi+\xi^\dagger \M\xi^\dagger)T]+\lambda_4\,
\tr(\ov{T}T)\tr(\M^\dagger\Sigma+\Sigma^\dagger\M).   \cr}\eqno(3.7)$$

   We begin by first examining the chiral-loop effects on the strong decays
$B_Q\ri B_Q'+\pi^a$. The tree amplitudes of $S_{ij}\ri S_{ij}+\pi^a$ and
$S_{ij}\ri T_{ij}+\pi^a$ derived from (3.4) are
$$\eqalign{
A[S_{ij}^\mu(v)\ri S_{ij}^\nu\pi^a(q)]= &\,i{3\over 8}\,{g_1\over
f_0}\,\ov{\U}^\nu\epsilon_{\nu\alpha\beta\mu}v^\alpha q^\beta(\lambda^a_{ii}
+\lambda^a_{jj}) \U^\mu,   \cr
A[S_{ij}^\mu(v)\ri T_{ij}\pi^a(q)]= &\,-{\sqrt{3}\over 2\sqrt{2}}\,{g_2\over
f_0}\,\bar{u}_{\bar{3}}q_\mu(\lambda^a_{ii}-\lambda^a_{jj})\U^\mu~~~(i<j),
\cr}\eqno(3.8)$$
with $\U_\mu=u_\mu-{1\over\sqrt{3}}(\gamma_\mu+v_\mu)\gamma_5u$. It is a
simple matter to project out from (3.8) the strong decay amplitudes
expressed in terms of component fields $B_6^*,~B_6,~B_{\bar{3}}$. For example,
Eq.(3.8) leads to
$$\eqalign{ A[(B_6^{*\mu})_{ij}\ri \,(B_6)_{ij}\pi^a(q)]=&\,{\sqrt{3}\over 8}
\,{g_1\over f_0}\,\bar{u}_{6} q_\mu (\lambda^a_{ii}+\lambda_{jj}^a)
u^\mu,    \cr
A[(B_6)_{ij}\ri (B_{\bar{3}})_{ij}
\pi^a(q)]= &\,{g_2\over 2\sqrt{2}f_0}\bar{u}_{{\bar{3}}} q\!\!\!/ \gamma_5
(\lambda^a_{ii}-\lambda^a_{jj})u_{6}~~~(i<j),   \cr}\eqno(3.9)$$
where use has been made of the identity (our convention being $\epsilon_{0123}
=1$)
$$i\epsilon^{\mu\nu\lambda\kappa}\gamma_\kappa\gamma_5=\,g^{\mu\nu}\gamma^
\lambda-g^{\mu\lambda}\gamma^\nu+g^{\nu\lambda}\gamma^\mu-\gamma^\mu\gamma
^\nu\gamma^\lambda,\eqno(3.10)$$
and $v\!\!\!/ u^\mu=u^\mu,~\gamma^\mu u_\mu=0,~v^\mu u_\mu=0$. One-loop
contributions are shown in Figs. 4 and 5 respectively for $S\ri S\pi$ and
$S\ri T\pi$. The resultant effective couplings are given by
$$(g_1)_{\rm eff}= \,g_1{Z_2(S)\sqrt{Z_2(\pi^a)}\over Z_1},\eqno(3.11)$$
and
$$(g_2)_{\rm eff}= \,g_2{\sqrt{Z_2(S)Z_2(T)Z_2(\pi^a)}\over Z_1'},\eqno(3.12)$$
where $Z_2(T)$ and $Z_2(S)$ are respectively the wave-function renormalization
constants for $T$ and $S$ baryon fields, $Z_1$ and $Z_1'$ account for vertex
renormalization effects from Figs. 4a-4d and Figs. 5a-5d, respectively.

To evaluate the renormalization constant $Z_2$, we note that the self-energy
amplitudes can be written as $-i\Sigma(\k)$ and $i\Sigma(\k)g_{\mu\nu}$
for $T$ and $S$ heavy baryon fields, respectively.
Since
$$\Sigma(\k)=\,\delta m-(Z_2^{-1}-1)v\cdot \k,\eqno(3.13)$$
we find
\foot{Our results for $Z_2(T)$ and $Z_2(S)$ [see also Eq.(3.18)] disagree with
Eq.(3.3) of Ref.[17].}
$$\eqalign{  \delta m(T_{ij})= &\,-{3\over 32\pi}\,g^2_2\sum_b\half\big[
(\lambda^b\lambda^b)_{ii}+(\lambda^b\lambda^b)_{jj}+2\lambda^b_{ij}\lambda^b_
{ji}-2\lambda^b_{ii}\lambda^b_{jj}\big]\,{\mpi^3\over f_{0}^2},  \cr
  Z_2(T_{ij})= &\,1+{9\over 2}\,g^2_2\sum_b\half\big[(\lambda^b\lambda^b)
_{ii}+(\lambda^b\lambda^b)_{jj}+2\lambda^b_{ij}\lambda^b_{ji}-2\lambda^b_{ii}
\lambda^b_{jj}\big]\,\epsilon_{\pi^b}~~(i\ne j),  \cr}
\eqno(3.14)$$
for the antitriplet baryon $B_{\bar{3}}$ with $\epsilon_{\pi^b}$'s being
defined in (2.22). Likewise, we get
$$\eqalign{ \delta m(S_{ij})=&\,-{1\over 128\pi}\sum_b(6g^2_1\xi^s_{ij}
+4g^2_2\zeta^s_{ij}){m^3_{\pi^b}\over f^2_{0}},    \cr Z_2(S_{ij})=&\,1+
{3\over 8}\sum_b \big(6g_1^2\xi^s_{ij}+4g^2_2\zeta^s
_{ij}\big)\epsilon_{\pi^b},   \cr}\eqno(3.15)$$
for the sextet baryons $B_6$ and $B_6^{*}$, where
$$\eqalign{ \xi^s_{ij}= &{1\over 4}\big[(\lambda^b\lambda^b)_{ii}+(\lambda^b
\lambda^b)_{jj}+2\lambda^b_{ij}\lambda^b_{ji}+2\lambda^b_{ii}\lambda^b_{jj}
(1-\delta_{ij})
\big],   \cr
\zeta^s_{ij}= &\half\big[(\lambda^b\lambda^b)_{ii}+(\lambda^b\lambda^b)
_{jj}-2\lambda^b_{ij}\lambda^b_{ji}-2\lambda^b_{ii}\lambda^b_{jj}(1-\delta_
{ij})\big]. \cr}\eqno(3.16)$$
To work out the wave function renormalization constant $Z_2$ for heavy
baryons, Eq.(2.32) and the following relations are useful:
$$\eqalign{
  & \sum_b(\lambda^b_{11}\lambda^b_{11})\epsilon_{\pi^b}=\sum_b(\lambda^b_{22}
\lambda^b_{22})\epsilon_{\pi^b}=\,\epsilon_\pi+{1\over 3}\epsilon_\eta,  \cr
  & \sum_b(\lambda^b_{13}\lambda^b_{31})\epsilon_{\pi^b}=\sum_b(\lambda^b_{23}
\lambda^b_{32})\epsilon_{\pi^b}=\,2\epsilon_{_K},  \cr
  & \sum_b(\lambda^b_{12}\lambda^b_{21})\epsilon_{\pi^b}=\,2\epsilon_\pi,~~~
\sum_b(\lambda^b_{33}\lambda^b_{33})\epsilon_{\pi^b}=\,{4\over 3}\epsilon_
\eta.  \cr}\eqno(3.17)$$
It follows from Eqs.(3.14) and (3.15) that
$$\eqalign{
  Z_2(\Lambda_Q)= &\,1+9g_2^2(3\epsilon_\pi+\epsilon_{_K}),  \cr
  Z_2(\Xi_Q)= &\,1+{9\over 4}g_2^2(3\epsilon_\pi+10\epsilon_{_K}+3\epsilon_
\eta),  \cr
  Z_2(\Sigma_Q)= &\,1+{9\over 8}g_1^2(4\epsilon_\pi+2\epsilon_{_K}+{2\over 3}
\epsilon_\eta)+3g_2^2(\epsilon_\pi+\epsilon_{_K}), \cr
  Z_2(\Xi'_Q)= &\,1+{9\over 8}g_1^2({3\over 2}\epsilon_\pi+5\epsilon_{_K}+
{1\over 6}\epsilon_\eta)+{3\over 4}g_2^2(3\epsilon_\pi+2\epsilon_{_K}+3
\epsilon_\eta),  \cr
 Z_2(\Omega_Q)= &\,1+{9\over 8}g_1^2(4\epsilon_{_K}+{8\over 3}
\epsilon_\eta)+6g_2^2\epsilon_{_K}.  \cr}\eqno(3.18)$$

   As for the vertex diagram Fig. 4a for $S_{ij}\ri S_{ij}+\vec{\pi}$
(since mass differences are generally small among
different spin multiplets of heavy baryons, we will thus only consider
pion emission due to the small available phase space), the general expression
is
$$Z_1(4a)=\,1+\sum_b\left({9\over 4}g_1^2\xi_{ij}-6g_2^2\zeta_{ij}\right)
{\epsilon_{\pi^b}\over \lambda^3_{ii}+\lambda^3_{jj} },\eqno(3.19)$$
with (see Appendix B for a derivation of the SU(3) group factors)
$$\eqalign{ \xi_{ij}= &\,{1\over 8}\bigg\{(\lambda^b\lambda^3
\lambda^b)_{jj}+\lambda^3_{ii}(\lambda^b\lambda^b)_{jj}+\lambda^b_{ji}(
\lambda^3\lambda^b)_{ij}+(\lambda^b\lambda^3)_{ji}\lambda^b_{ij}   \cr
&+[\,\lambda^3_{ij}(\lambda^b\lambda^b)_{ji}+(\lambda^b\lambda^3)_{jj}\lambda^b
_{ii}+\lambda^b_{jj}(\lambda^3\lambda^b)_{ii}\,](1-\delta_{ij})+(i
\leftrightarrow j)\bigg\},  \cr
\zeta_{ij}= &\,{1\over 4}\bigg\{(\lambda^b\lambda^3
\lambda^b)_{jj}-\lambda^3_{ii}(\lambda^b\lambda^b)_{jj}
-\lambda^3_{ij}(\lambda^b\lambda^b)_{ji}(1-\delta_{ij})+(i
\leftrightarrow j)\bigg\}.  \cr}\eqno(3.20)$$
Note that $Z_1$ in Eq.(3.19) is worked out for $\pi^0$ emission, but due to
SU(2) symmetry it should be also valid for charged pion emission.
We next express the vertex corrections explicitly
for $S(\Sigma_Q^{(*)})\ri S(\Sigma_Q)\vec{\pi}$ and $S({\Xi'}_Q^{(*)})\ri
S({\Xi'}_Q)\vec{\pi}$:
$$Z_1(4a)=1+\big({9\over 8}g^2_1+6g_2^2\big)\big(\epsilon_\pi
+\half\epsilon_{_K}\big)+{3\over 8}g_1^2\epsilon_\eta\eqno(3.21)$$
for $S(\Sigma_Q^{(*)})\ri S(\Sigma_Q)\vec{\pi}$, and
$$ Z_1(4a)= \,1+{9\over 8}\,g^2_1\left(-{1\over 4}
\epsilon_\pi+2\epsilon_{_K}+{1\over 12}\epsilon_\eta
\right)  +6g_2^2\left({1\over 4}\epsilon_\pi+\epsilon_{_K}+{1\over 4}
\epsilon_\eta\right)\eqno(3.22)$$
for $S({\Xi'}_Q^{(*)})\ri S({\Xi'}_Q)\vec{\pi}$. Similarly, for $S_{ij}
\ri T_{ij}+\vec{\pi}$, Fig. 5a leads to
$$Z_1'(5a)=\,1+\sum_b\left(-{9\over 2\sqrt{2}}g_1^2\xi_{ij}'+{3 \over
\sqrt{2}}g_2^2\zeta_{ij}'
\right){\epsilon_{\pi^b}\over \lambda^3_{ii}-\lambda^3_{jj}}~~~(i<j),
\eqno(3.23)$$
with
$$\eqalign{ \xi_{ij}'= &\,{1\over 4 \sqrt{2}}\bigg\{(\lambda^b\lambda^3
\lambda^b)_{ii}+\lambda^3_{jj}(\lambda^b\lambda^b)_{ii}+\lambda^b_{ij}
(\lambda^3\lambda^b)_{ji}+(\lambda^b\lambda^3)_{ij}\lambda^b_{ji}
-(i\leftrightarrow j)\bigg\}, \cr
\zeta_{ij}'= &\,{1\over 2 \sqrt{2}}\bigg\{(\lambda^b\lambda^3
\lambda^b)_{ii}-\lambda^3_{jj}(\lambda^b\lambda^b)_{ii}
+\lambda^b_{ii}(\lambda^3\lambda^b)_{jj}-(\lambda^b\lambda^3)
_{ii}\lambda^b_{jj}-(i\leftrightarrow j)\bigg\}.  \cr}\eqno(3.24)$$
Explicitly,
$$Z_1'(5a)=1+{9\over 2}g_1^2\left(\epsilon_\pi+{1\over 4}\epsilon
_{_K}\right)+3g^2_2\left(\epsilon_\pi+\half\epsilon_{_K}\right)\eqno(3.25)$$
for $S(\Sigma_Q^{(*)})\ri T(\Lambda_Q)\vec{\pi}$ and
$$Z_1'(5a)= \,1+{9\over 2}g_1^2\left({1\over 8}\epsilon
_\pi+\epsilon_{_K}+{1\over 8}\epsilon_\eta
\right) +3g_2^2\left(-{1\over 4}\epsilon_\pi+\epsilon_{_K}
+{3\over 4}\epsilon_\eta\right)\eqno(3.26)$$
for $S({\Xi'}_Q^{(*)})\ri T(\Xi_Q)\vec{\pi}$.

   The cancellation between the wave-function renormalization of the
Goldstone bosons and the 5-point vertex diagram for heavy-meson strong decays
also persists in the baryon sector. This can be understood from the fact that
the 5-point vertex in both cases arises from expanding the chiral field
${\cal A}_\mu$ to the third power in $M$ and hence has the same structure.
Therefore, $Z_1(4d)=Z'_1(5d)=\sqrt{Z_2(\pi^a)}$.
Just like their counterparts in the meson sector,  all the seagull diagrams in
Figs. 4 and 5 also do not contribute to the strong decays $S\ri S\pi$ and $S
\ri T\pi$ to the lowest order in heavy quark expansion. First of all,
we see from Eqs.(3.4) and (2.24) that the four-point vertices
$SS\pi^a\pi^b$ and $TT\pi^a\pi^b$ are proportional to $(q_a-q_b)\cdot v$,
where $q_a$ is the 4-momentum of the $\pi^a$. Second, there is no $q$
dependence in the denominators of the amplitudes
for Figs. 4c and 5c as the loop momenta can
be chosen to be $l$ and $l+\tilde{k}$ for the light meson and heavy baryon,
respectively. As for Figs. 4b and 5b, the $q$ dependent terms in both numerator
and denominator are always of the form $(q\cdot v)$. As a consequence, it will
not give rise to the desired amplitudes shown in Eq.(3.8) after loop
integration. We thus conclude that
$$(g_1)_{\rm eff}= \,g_1{Z_2(S)\over Z_1(4a)},\eqno(3.27)$$
and
$$(g_2)_{\rm eff}= \,g_2{\sqrt{Z_2(S)Z_2(T)}\over Z_1'(5a)}.\eqno(3.28)$$

   We now switch to the electromagnetic (M1) decays $S\ri S+\gamma$
and $S\ri T+\gamma$. The most general gauge invariant Lagrangian for magnetic
transitions of heavy baryons given in paper II (II.3.9) can be recast in
terms of superfields
\foot{The sign of the last two terms in Eq.(3.29) is opposite to those of the
corresponding terms in Eq.(18) of Ref.[7].}
$$\eqalign{ \L_B^{(2)}= &\,-i3a_1\,\tr(\bar{S}^\mu\Q F_{\mu\nu}S^\nu)+
\sqrt{3}a_2\epsilon_{\mu
\nu\alpha\beta}\,\tr(\bar{S}^\mu\Q v^\nu F^{\alpha\beta}T)+h.c.  \cr
& +3a_1'\,\tr(\bar{S}^\mu\Q' \sigma\cdot FS_\mu)
-{3\over 2}a_1'\tr(\ov{T}\Q'\sigma\cdot FT),  \cr}\eqno(3.29)$$
where $\sigma\cdot F\equiv\sigma_{\mu\nu}F^{\mu\nu}$,
and we have applied Eqs.(II.3.47) and (II.3.61).
The Lagrangian ${\cal L}_B^{(2)}$ is also the most general chiral-invariant
   one provided that one makes the replacement
$${\cal Q}\rightarrow {1\over 2}(\xi^{\dagger}{\cal Q}\xi+\xi {\cal Q}
\xi^{\dagger}), \ \ \ {\cal Q^{\prime}}\rightarrow {\cal Q^{\prime}}.
\eqno(3.30)$$
Note that contrary to Eq.(II.3.9), the Dirac magnetic moment terms do not
appear in (3.29) because the Lagrangian $\L_B^{(1)}$ is expressed in
terms of velocity dependent fields and hence does not contain Dirac magnetic
moments to the lowest order. In the quark model the
magnetic moments $a_1$ and $a_2$ are simply related to the Dirac magnetic
moments of the light quarks whereas $a_1^{\prime}$ is connected
to those of heavy quarks. Explicitly, $a_1'$ is fixed by heavy quark symmetry
to be ${1\over 6}{e\over 2m_Q}$ and this is the only $1/m_Q$ effect included
for heavy baryon decays. In contrast, the couplings $a_1$ and $a_2$ are
independent of the heavy quark mass and spin.

    It follows from (3.29) that the tree amplitudes of heavy baryon
radiative decays are  given by
$$\eqalign{ A[S_{ij}^\mu(v)\ri S_{ij}^\nu+\gamma(\ep,k)]= &i\,{3\over 2}
a_1\,\ov{\U}^\nu(\Q_{ii}+\Q_{jj})(k_\nu\ep_\mu-k_\mu\ep_\nu)\,\U^
\mu-i6a_1'\Q'\,\ov{\U}^\mu k\!\!\!/\ep\!\!/\U_\mu,  \cr
A[S_{ij}^\mu(v)\ri T_{ij}+\gamma(\ep,k)]= &\,-2\sqrt{3\over 2}
a_2\,\epsilon_{\mu\nu\alpha\beta}\,\bar{u}_{\bar{3}}v^\nu k^\alpha\ep^
\beta(\Q_{ii}-\Q_{jj})\,\U^\mu~~~(i<j).    \cr}\eqno(3.31)$$
For the radiative decay $S\ri T+\gamma$, the two
light quarks in the heavy hadron must undergo a spin-flip transition.
Consequently, such decays will not receive any contributions from the magnetic
moment of the heavy quark. As we will see later, this property persists at the
one-loop level. The decay amplitudes in terms of component
fields can be projected out from (3.31) by using the following relations:
$$\eqalign{  \epsilon_{\mu\nu\alpha\beta}\bar{u}v^\nu k^\alpha\ep^\beta
\U^\mu= &\,i\bar{u}(k_\mu\ep\!\!/-\ep_\mu k\!\!\!/ )\gamma_5
u^\mu+{1\over\sqrt{3}}\bar{u}\sigma_{\mu\nu}k^\mu\ep^\nu u,   \cr
\ov{\U}^\nu(k_\nu\ep_\mu-k_\mu\ep_\nu)\U^\mu= &\,\bar{u}^\nu(k_\nu\ep_\mu
-k_\mu\ep_\nu)u^\mu+{1\over\sqrt{3}}\bar{u}(k_\mu\ep\!\!/
-\ep_\mu k\!\!\!/)\gamma_5u^\mu  \cr
-&\, {1\over\sqrt{3}}\bar{u}^\mu(k_\mu\ep\!\!/-\ep_\mu k\!\!\!/)\gamma_5u
+ i{2\over 3}\bar{u}\sigma_{\mu\nu}k^\mu\ep^\nu u,  \cr
\ov{\U}^\mu k\!\!\!/\ep\!\!/\U_\mu=&\,\bar{u}^\mu k\!\!\!/\ep\!\!/u_
\mu-{2\over\sqrt{3}}\bar{u}(k_\mu\ep\!\!/-\ep_\mu k\!\!\!/)\gamma_5
u^\mu  \cr + &\,{2\over\sqrt{3}}\bar{u}^\mu(k_\mu\ep\!\!/-\ep_\mu k\!\!\!/)
\gamma_5u
- \,i{1\over 3}\bar{u}\sigma_{\mu\nu}k^\mu\ep^\nu u.  \cr}\eqno(3.32)$$
To the lowest order, we have the following predictions from (3.31):
$$\eqalign{  A({\Xi'}_Q^\half\ri \Xi_Q^\half\gamma)=&\,A(\Sigma_Q^0\ri\Lambda
_Q\gamma)=\,-\sqrt{2}a_2\,\bar{u}\sigma_{\mu\nu}k^\mu\ep^\nu u,   \cr
A({\Xi'}_Q^{-\half}\ri\Xi_Q^{-\half}\gamma)= &\,A({\Xi'}_Q^{*-\half}\to \Xi_Q^
{-\half}\gamma)=\,0, \cr  A({\Xi'}_Q^{*\half}\ri \Xi_Q^\half\gamma)=&\,A(
\Sigma_Q^{*0}\ri\Lambda_Q\gamma)=\,-i\sqrt{6}a_2\,\bar{u}(k_\mu\ep\!\!/-\ep_
\mu k\!\!\!/)\gamma_5u^\mu,   \cr   A({\Xi'}_Q^{*-\half}\ri{\Xi'}_Q^{-\half}
\gamma)= &\,A(\Sigma^{*-1}_Q\ri\Sigma_Q^{-1}\gamma)=\,A(\Omega_Q^{*}\ri
\Omega_Q\gamma) \cr  = &\,-{i\over\sqrt{3}}(a_1-8a_1')\,\bar{u}(k_\mu
\ep\!\!/ -\ep_\mu k\!\!\!/)\gamma_5u^\mu,   \cr
A({\Xi'}_Q^{*\half}\ri{\Xi'}_Q^\half\gamma)= &\,A(\Sigma^{*0}
_Q\ri\Sigma_Q^0\gamma)=\,{i\over 2\sqrt{3}}(a_1+16a_1')\,\bar{u}(k_\mu
\ep\!\!/-\ep_\mu k\!\!\!/)\gamma_5u^\mu,   \cr}\eqno(3.33)$$
where the superscript denotes the isospin quantum number.
The above SU(3) relations will be violated in the presence of chiral-loop
contributions as depicted in Figs. 6 and 7. Chiral corrections will modify the
coupling constants $a_1,~a_1'$ and $a_2$ to
$$\eqalign{ (a_1)_{\rm eff}= &\,a_1{Z_2(S)\over Z_1}+\delta a_1(6c+6d+6e), \cr
 (a_1')_{\rm eff}(S)= &\,a_1'{Z_2(S)\over Z_1'(SS\gamma)}=\,a'_1,~~~~~
 (a_1')_{\rm eff}(T)=\,a_1'{Z_2(T)\over Z_1'(TT\gamma)}=\,a'_1,  \cr
(a_2)_{\rm eff}= &\,a_2{\sqrt{Z_2(S)Z_2(T)}\over Z_1''}+\delta a_2(7c+7d+7e),
 \cr}\eqno(3.34)$$
where $Z_1\,(Z_1'(SS\gamma))$, $Z_1''$ and $Z'_1(TT\gamma)$
are the vertex renormalization constants induced by Figs. 6a-6b, 7a-7b and 8,
respectively. In (3.34) we have anticipated the results [see (3.38) and (3.40)
below]
$$Z_2(S)=\,Z'_1(SS\gamma),~~~Z_2(T)=\,Z'_1(TT\gamma).\eqno(3.35)$$
We remark
that since ${\Xi'}_Q^{-\half(*)}\ri\Xi_Q^{-\half}\gamma$ is prohibited at
tree level, its effective coupling $(a_2)_{\rm eff}$ is defined in a similar
way as ${\Xi'}_Q^{\half(*)}\ri\Xi_Q^\half\gamma$:
$$A[S({\Xi'}_Q^{-\half(*)})\ri T(\Xi_Q^{-\half})\gamma]=\,-2\sqrt{3\over 2}
(a_2)_{\rm eff}\,\bar{u}\epsilon_{\mu\nu\alpha\beta}v^\nu k^\alpha\ep
^\beta{\U}^\mu,\eqno(3.36)$$
with
$$(a_2)_{\rm eff}=\,\delta a_2(7a+7b+7c+7d+7e).\eqno(3.37)$$

Having calculated the renormalization constants $Z_2(S)$ and $Z_2(T)$ earlier,
we now proceed to evaluate Figs. 6a-6b, 7a-7b and 8 for the vertex
renormalization constants $Z_1,~Z'_1,~Z_1''$. Note
that charge couplings for the vertices $SS\gamma$ and $TT\gamma$
in Fig. 6a and $SS\gamma$ in Fig. 7a are
proportional to $(v\cdot\epsilon)$ and consequently cannot induce magnetic
transitions. The contributions to $Z_1$ due to the magnetic
$SS\gamma$ and $ST\gamma$ couplings can be summarized as follows:
$$\eqalign{ Z_1(SS\gamma;6a)= &\,1+\sum_b\left({9\over 4}g_1^2\tilde{\xi}_{
ij}-6{a_2\over a_1}g_1g_2\tilde{\zeta}_{ij}\right){\epsilon_{\pi^b}\over
\Q_{ii}+\Q_{jj}},   \cr  Z_1'(SS\gamma;6a)= &\,1+\sum_b{3\over 2}\left({3\over
2}g_1^2\tilde{\xi}_{ij}'+{1 \over 2}
g_2^2\tilde{\zeta}'_{ij}\right)\epsilon_{\pi^b}, \cr}\eqno(3.38)$$
with
$$\eqalign{ \tilde{\xi}_{ij}= &\,{1\over 8}\bigg\{(\lambda^b\Q\lambda^b)
_{jj}+(\lambda^b\lambda^b)_{jj}\Q_{ii}+2\lambda^b_{ji}\lambda^b_{ij}\Q_{ii}
+\lambda^b_{jj}\lambda^b_{ii}(\Q_{ii}+\Q_{jj})(1-\delta_{ij})+(i
\leftrightarrow j)\bigg\},  \cr
\tilde{\zeta}_{ij}= &\,{1\over 4}\left\{(\lambda^b\Q\lambda^b)_{jj}-(\lambda
^b\lambda^b)_{jj}\Q_{ii}+(i\leftrightarrow j)\right\},   \cr
\tilde{\xi}_{ij}'=&\,{1\over 4}\left\{(\lambda^b\lambda^b)_{jj}+(\lambda^b
\lambda^b)_{ii}+2\lambda^b_{ji}\lambda^b_{ij}+2\lambda^b_{jj}\lambda^b_{ii}
(1-\delta_{ij})\right\},  \cr
\tilde{\zeta}_{ij}'=&\,\left\{(\lambda^b\lambda^b)_{jj}+(\lambda^b
\lambda^b)_{ii}-2\lambda^b_{ji}\lambda^b_{ij}-2\lambda^b_{jj}\lambda^b_{ii}
(1-\delta_{ij})\right\}, \cr}\eqno(3.39)$$
for $S_{ij}\ri S_{ij}+\gamma$,
$$Z'_1(TT\gamma;8)=\,1+{9\over 4}\sum_b g_2^2\left[(\lambda^b\lambda^b)_{ii}+
(\lambda^b
\lambda^b)_{jj}+2\lambda^b_{ij}\lambda^b_{ji}-2\lambda^b_{ii}\lambda^b_{jj}
\right]\epsilon_{\pi^b}~~(i< j)\eqno(3.40)$$
for $T_{ij}\ri T_{ij}+\gamma$,
\foot{Since the magnetic $TT\gamma$ coupling vanishes in heavy quark limit
and chiral corrections preserve heavy quark symmetry, we can be sure that
no other diagrams such as the seagull diagrams and the diagram like Fig. 8 but
with a photon attached to the light meson can contribute to $Z_1(T\ri
T\gamma)$. We have checked this explicitly.}
and
$$\eqalign{ Z_1''(ST\gamma;7a)= &\,1+\sum_b\left(-{9\sqrt{2}\over 4}g_1g_2{a_1
\over a_2}{\tilde{\xi}_{ij}''\over \Q_{ii}-\Q_{jj}}+{3\over\sqrt{2}}g_2^2{
\tilde{\zeta}_{ij}''\over \Q_{ii}-\Q_{jj}}\right)\epsilon_{\pi^b} ~~~(i<j),
\cr}  \eqno(3.41)$$
with
$$\eqalign{ \tilde{\xi}_{ij}''= &\,{1\over \sqrt{2}4}[\,(\lambda^b\Q\lambda^b)
_{ii}+(\lambda^b\lambda^b)_{ii}\Q_{jj}+2\lambda^b_{ij}\lambda^b_{ji}\Q_{jj}
-(i\leftrightarrow j)\,],   \cr
\tilde{\zeta}_{ij}''= &\,{1\over 2\sqrt{2}}[\,(\lambda^b\Q\lambda^b)
_{ii}-(\lambda^b\lambda^b)_{ii}\Q_{jj}+\lambda^b_{ii}\lambda^b_{jj}(\Q_{jj}
-\Q_{ii})-(i\leftrightarrow j)\,],   \cr}\eqno(3.42)$$
for $S_{ij}\ri T_{ij}+\gamma$. It is easily seen from Eqs.(3.38), (3.40) and
(3.14-3.15) that (3.35) is confirmed; the exact QCD result that $a'_1$ does
not get renormalized [13] is thus realized in chiral perturbation theory.
Note that when $\Q_{ii}=\Q_{jj}$, the tree
amplitude for $S\ri T+\gamma$ vanishes [see Eq.(3.31)] and $Z_1''$ is
undefined. In this case, one should apply Eqs.(3.36) and (3.37).
The explicit expressions of the vertex renormalization constant $Z_1$
for each radiative decay mode are then given by
$$\eqalign{
Z_1(\Sigma_Q^{+1(*)}\ri\Sigma_Q^{+1}\,\gamma;6a)=&\,1+{9\over 8}g_1^2
({5\over 4}\epsilon_\pi+{1\over 4}\epsilon_{_K}+{1\over 3}\epsilon_\eta)
+{9\over 2}g_1g_2\,{a_2\over a_1}(\epsilon_\pi+\epsilon_{_K}),  \cr
Z_1(\Sigma_Q^{0(*)}\ri\Sigma_Q^0\,\gamma;6a)=&\,1+{9\over 4}g_1^2
(\epsilon_\pi-{1\over 4}\epsilon_{_K}+{1\over 6}\epsilon_\eta)
+9g_1g_2\,{a_2\over a_1}(\epsilon_{_K}),  \cr
Z_1(\Sigma_Q^{-1(*)}\ri\Sigma_Q^{-1}\,\gamma;6a)=&\,1+{9\over 8}g_1^2
({1\over 2}\epsilon_\pi+\epsilon_{_K}+{1\over 3}\epsilon_\eta)
+9g_1g_2\,{a_2\over a_1}(\epsilon_\pi),  \cr
Z_1({\Xi'}_Q^{\half(*)}\ri{\Xi'}_Q^{\half}\,\gamma;6a)=&\,1+{27
\over 16}g_1^2 (-{1\over 2}\epsilon_\pi+{5\over 3}\epsilon_{_K}+{1\over 18}
\epsilon_\eta)+{9\over 2}g_1g_2{a_2\over a_1}(-\epsilon_\pi+2\epsilon_{_K}+
\epsilon_\eta),  \cr
Z_1({\Xi'}_Q^{-\half(*)}\ri{\Xi'}_Q^{-\half}\,\gamma;6a)=&\,1+{9
\over 32}g_1^2(7\epsilon_{_K}+{1\over 3}\epsilon_\eta)
+{9\over 2}g_1g_2\,{a_2\over a_1}(\epsilon_\pi+\epsilon_{_K}),  \cr
Z_1(\Omega_Q^{(*)}\ri\Omega_Q\,\gamma;6a)=&\,1+{3\over 4}g_1^2
({3\over 4}\epsilon_{_K}+2\epsilon_\eta)
+9g_1g_2\,{a_2\over a_1}(\epsilon_{_K}),  \cr}\eqno(3.43)$$
and
$$\eqalign{
Z_1''(\Sigma_Q^{0(*)}\ri\Lambda_Q\,\gamma;7a)= &\,1+{9\over 2}g_1g_2{a_1\over
a_2}(\epsilon_\pi+{1\over 4}\epsilon_{_K})+3g^2_2(\epsilon_\pi+\half\epsilon_
{_K}),   \cr
Z_1''({\Xi_Q'}^{\half(*)}\ri\Xi_Q^\half\,\gamma;7a)= &\,1+{9\over 8}g_1g_2
{a_1\over a_2}(\half\epsilon_\pi+{13\over 3}\epsilon_{_K}+{1\over 6}\epsilon_
\eta)
+{3\over 4}g^2_2(\epsilon_\pi+2\epsilon_{_K}+3\epsilon_\eta),  \cr
\delta a_2({\Xi'}_Q^{-\half(*)}\ri\Xi_Q^{-\half}\,\gamma;7a)= &\,{3\over 8}
g_1g_2a_1(-\epsilon_{_K}+\epsilon_\eta)+{3\over 2}g^2_2a_2(-\epsilon_\pi+
\epsilon_{_K})    \cr}\eqno(3.44)$$
for $S\to T+\gamma$. Recall that the heavy quark magnetic moment
$a'_1$ does not contribute to the $S\ri T\gamma$ transitions at the tree
level. Here, we see that this consequence of the heavy quark symmetry is
preserved at the loop level.

   The electromagnetic five-point vertices in Figs. 6b and 7b are generated by
the Lagrangian (3.29) with the replacement $\Q \ri {1\over 4f_0^2}\,[M,~[\Q,
{}~M]]$ which comes from (3.30). The results are
$$\eqalign{ Z_1(S_{ij}\ri S_{ij}+\gamma;6b)=&\,1+{1\over 4}\sum_b\Big\{[
\lambda^b,~[\Q,~\lambda^b]]_{ii}+(i\ri j)\Big\}{\epsilon_{\pi^b}\over
\Q_{ii}+\Q_{jj}},  \cr
Z_1''(S_{ij}\ri T_{ij}+\gamma;7b)= &\,1+{1\over 4}\sum_b\Big\{[\lambda^b,~
[\Q,~\lambda^b]]_{ii}-(i\ri j)\Big\}{\epsilon_{\pi^b}\over \Q_{ii}-\Q_{jj}}
{}~~(i<j).  \cr}\eqno(3.45)$$
Explicit expressions for $Z_1(6b)$ and $Z_1''(7b)$ for individual decay modes
read
$$\eqalign{ Z_1(\Sigma^{+1(*)}_Q\ri\Sigma_Q^{+1}\,\gamma;6b) =&\,
Z_1({\Xi'}_Q^{-\half(*)}\ri{\Xi'}_Q^{-\half}\,\gamma;6b)=\,
1-{3\over 2}(\epsilon_\pi+\epsilon_\K),  \cr
  Z_1(\Omega_Q^{(*)}\ri\Omega_Q\,\gamma;6b)= &\,
 Z_1(\Sigma^{0(*)}_Q\ri\Sigma_Q^0\,\gamma;6b) =\,1-3\epsilon_\K,  \cr
  Z_1(\Sigma_Q^{-1(*)}\ri\Sigma_Q^{-1}\,\gamma;6b) =&\,
 Z_1({\Xi'}_Q^{\half(*)}\ri{\Xi'}_Q^{\half}\,\gamma;6b) =\,1-3\epsilon_\pi,
\cr}\eqno(3.46)$$
and
$$\eqalign{  Z''_1(\Sigma_Q^{0(*)}\ri\Lambda_Q\,\gamma;7b)= &\,1-(2\epsilon_
\pi+\epsilon_\K),  \cr
Z''_1({\Xi'}_Q^{\half(*)}\ri{\Xi}_Q^\half\,\gamma;7b) =&\,
1-(\epsilon_\pi+2\epsilon_\K),   \cr
\delta a_2({\Xi'}_Q^{-\half(*)}\ri\Xi_Q^{-\half}\,\gamma;7b) =&\,
(-\epsilon_\pi+\epsilon_\K)a_2.  \cr}\eqno(3.47)$$

   Figs. 6c and 7c are the Feynman diagrams in which a photon couples to
light pseudoscalar mesons. Care must be taken in order to ensure that the
radiative decay amplitude is gauge invariant. As an example, we consider
the decay $\Sigma_Q^{*+1}\to \Sigma_Q^{+1}\gamma$ through the charged pion
loop and the sextet baryon intermediate state. Its amplitude is given by
$$A_\pi(S;6c)=\,-{9\over 8}\,{g_1^2\over f_0^2}\,\ov{\U}^\nu\int {d^4l\over
(2\pi)^4}\,{\epsilon_{\nu
\rho\sigma\beta}\epsilon^{\beta\lambda\delta\mu}v_\lambda l_\delta v^\rho(l+k)
^\sigma(l\cdot\ep)\over [(l+k)^2-m^2_\pi+i\epsilon](l^2-m^2_\pi+i\epsilon)
(v\cdot l+i\epsilon) }\,\U_\mu,\eqno(3.48)$$
where we have kept the superfield notation.
We see that, aside from the $k^\sigma$ term in the numerator,
the ($l\cdot k$) term arising from the first denominator
in (3.48) also contributes; that is, they are all necessary for the sake
of gauge invariance. By applying Eqs.(A3) and (A5) we arrive at
$$A_\pi(S;6c)=\,i{9\over 256}\,{m_\pi\over \pi}\,{g_1^2\over f_0^2}\,\ov{\U}
^\nu(k_\nu\ep_\mu-k_\mu\ep_\nu)\,\U^\mu.\eqno(3.49)$$
Likewise, the amplitude for the charged pion loop with antitriplet baryon
intermediate state reads
$$A_\pi(T;6c)=\,i{3\over 32}\,{m_\pi\over \pi}\,{g_2^2\over f_0^2}\,\ov{\U}
^\nu(k_\nu\ep_\mu-k_\mu\ep_\nu)\,\U^\mu.\eqno(3.50)$$
Therefore, the radiative decay amplitude
due to the sextet or antitriplet intermediate baryon state is separately
gauge invariant, as it should be.
Projecting out  the spin-$\half$ final state from Eqs.(3.49)
and (3.50) gives
$$A_\pi(\Sigma_Q^{*+1}\to\Sigma_Q^{+1}\gamma;6c)=\,i{\sqrt{3}\over 32}\,{m_\pi
\over \pi f_0^2}\,({3\over 8}g_1^2+g_2^2)\bar{u}
(k_\mu\ep\!\!/ -\ep_\mu k\!\!\!/ )\gamma_5u^\mu.\eqno(3.51)$$
Comparing this with the tree amplitude
$$A(\Sigma_Q^{*+1}\to\Sigma_Q^{+1}\gamma)_{\rm tree}=\,i{2\over\sqrt{3}}(a_1
+6a_1'e_Q)
\bar{u}(k_\mu\ep\!\!/-\ep_\mu k\!\!\!/)\gamma_5u^\mu\eqno(3.52)$$
leads to
$$\delta a_1(\Sigma_Q^{*+1}\to\Sigma_Q^{+1}\gamma;6c)=\,{3e\over 64\pi}({3
\over 8}g_1^2+g_2^2){m_\pi\over f_0^2}\eqno(3.53)$$
for charged pion loops. On the contrary, the aforementioned ($l\cdot k$) term
does not contribute to the decay $S\ri T+\gamma$ due to the presence of
 the totally antisymmetric tensor $\epsilon_{\mu\nu\alpha\beta}$.
The general results including all charged meson loops are (see Appendix B for
a derivation of the SU(3) group factors)
$$\eqalign{ \delta a_1(6c)= &\,\sum_b{e\over 32\pi}{\mpi\over f_{0}^2}
\left({3\over 4}g_1^2\bar{\xi}_{ij}+g_2^2\bar{\zeta}_{ij}\right){1\over
\Q_{ii}+\Q_{jj}},  \cr \delta a_2(7c) =&\,\sum_b{3e\over 64\sqrt{2}\pi}g_1g_2
\,{\mpi\over f_{0}^2}\,{\bar{\xi}_{ij}'\over \Q_{ii}-\Q_{jj}}~~~(i<j),
 \cr}\eqno(3.54)$$
with
$$\eqalign{ \bar{\xi}_{ij}= &\,-{1\over 4}\bigg\{([\lambda^b,~\Q]\lambda^b)_
{ii}+(i\leftrightarrow j)\bigg\},   \cr
 \bar{\zeta}_{ij}= &\,-{1\over 2}\bigg\{([\lambda^b,~\Q]\lambda^b)_
{ii}+(i\leftrightarrow j)\bigg\},   \cr
 \bar{\xi}_{ij}'= &\,-{1\over 2\sqrt{2}}\bigg\{([\lambda^b,~\Q]\lambda^b)_
{ii}+[\lambda^b,~\Q]_{ij}\lambda^b_{ji}
-(i\leftrightarrow j)\bigg\}.  \cr}  \eqno(3.55)$$
Working out the above general expressions, we obtain
$$\eqalign{ \delta a_1(\Sigma^{+1(*)}_Q\ri\Sigma_Q^{+1}\,\gamma;6c) =&\,
\delta a_1({\Xi'}_Q^{-\half(*)}\ri{\Xi'}_Q^{-\half}\,\gamma;6c)  \cr =&\,
{3e\over 64\pi}\left({3\over 8}g_1^2+g_2^2\right)\left({m_\pi\over f_0^2}+{m_
{_K}\over f_{0}^2}\right),   \cr
  \delta a_1(\Omega_Q^{(*)}\ri\Omega_Q\,\gamma;6c)= &\,
 \delta a_1(\Sigma^{0(*)}_Q\ri\Sigma_Q^{0}\,\gamma;6c) \cr =&\,
{3e\over 32\pi}\left({3\over 8}g_1^2+g_2^2\right)\left({m_{_K}\over f_{0}^2}
\right),   \cr
  \delta a_1(\Sigma_Q^{-1(*)}\ri\Sigma_Q^{-1}\,\gamma;6c) =&\,
\delta a_1({\Xi'}_Q^{\half(*)}\ri{\Xi'}_Q^{\half}\,\gamma;6c)  \cr =&\,
{3e\over 32\pi}\left({3\over 8}g_1^2+g_2^2\right)\left({m_\pi\over f_0^2}
\right),  \cr }\eqno(3.56)$$
and
$$\eqalign{   \delta a_2(\Sigma_Q^{0(*)}\ri\Lambda_Q\,\gamma;7c)
=&\,{3e\over 32}\,{g_1g_2\over
\pi}\left({m_\pi\over f_0^2}+{1\over 4}\,{m_{_K}\over f_{0}^2}\right),  \cr
\delta a_2({\Xi'}_Q^{\half(*)}\ri{\Xi}_Q^\half\,\gamma;7c) =&\,{3e\over 32}\,
{g_1g_2\over
\pi}\left({1\over 4}{m_\pi\over f_0^2}+{m_{_K}\over f_{0}^2}\right),  \cr
\delta a_2({\Xi'}_Q^{-\half(*)}\ri\Xi_Q^{-\half}\,\gamma;7c) =&\,{3e\over 128}
\,{g_1g_2\over
\pi}\left(-{m_\pi\over f_0^2}+{m_{_K}\over f_{0}^2}\right).  \cr}
\eqno(3.57)$$
Note that $\delta a_2$ for the decay ${\Xi'}_Q^{-\half(*)}\ri\Xi_Q^{-\half}$
is defined in Eq.(3.37) and it does not contain the factor of $1/(\Q_{ii}-\Q_
{jj})$.
We leave it to the reader to check that, in the SU(3) limit, all above results
(3.43)-(3.57) do satisfy the SU(3) relations given by (3.33).

It remains to see if the seagull diagrams are important. The relevant
electromagnetic four-point vertices are described by
$$-{3\over 2\sqrt{2}}\,{eg_1\over f_0}\epsilon_{\mu\alpha\beta\nu}\bar{S}
^\mu v^\alpha A^\beta[\Q,~M]S^\nu+i\sqrt{3\over 2}{eg_2\over f_0}\bar{S}
^\mu A_\mu[\Q,~M]T,\eqno(3.58)$$
where uses of Eqs.(3.4) and (2.49) have been made. Evidently, the vertices $SS
\pi^a\gamma$ and $ST\pi^a\gamma$ are independent of the photon's momentum $k$.
This feature allows us to
conclude immediately that Figs. 6e and 7e cannot cause magnetic
transitions for the same reason as mentioned before for the case of
$S\ri S(T)\gamma$. The remaining two seagull diagrams Figs. 6d and 7d also
can be shown vanishing because of the identity $v_\mu u^\mu=0$. Hence,
$$\delta a_1(6d+6e)=\,\delta a_2(7d+7e)=0.\eqno(3.59)$$

   Finally, we note that the vertices $SS\pi^a$ and $ST\pi^a$ in the
Lagrangian (3.4) may suggest that there is a possibility of mixing between
the sextet and antitriplet baryons through loop effects. However, this is
not the case. An explicit calculation shows that such loop diagrams vanish.
The only possible mixing between $S$ and $T$ requires an $\epsilon_{\mu\nu
\alpha\beta}$ tensor, but there are not enough variables to construct
a Lorentz invariant.

   In summary, taking into account the mass differences among the Goldstone
bosons, we found that SU(3) relations are generally broken in the
strong and radiative decays of heavy hadrons. The leading chiral-loop
corrections have nonanalytic dependence on $m_q$ of the form
$m_q^\half$ or $m_q\ln m_q$.

\vskip 1.0cm

\noindent{\bf 4.~~~Applications}
\vskip 0.3 cm
    In Sections 2 and 3 we have presented a general analysis of
SU(3)-breaking effects in chiral perturbation theory for
 the strong and M1 radiative decays
of heavy mesons and heavy baryons. In this section we will apply the results
obtained so far to some selected decay modes of charmed mesons and baryons.
   Specifically, we choose the radiative decays $D^*\ri D\gamma,~\Sigma_c\ri
\Lambda_c\gamma,~\Xi_c'\ri\Xi_c\gamma,~{\Xi'}_c^*\ri\Xi_c'\gamma,~\Sigma_c
^*\to\Sigma_c\gamma,~\Omega_c^*\ri\Omega_c\gamma$
 and the strong decays $D^*\ri D\pi$,
$\Sigma_c\ri\Lambda_c\pi$ as examples of applications. The decay rates of
many of these modes have been explicitly calculated in papers I and II.
Special attention is
paid to the SU(3) relations (2.45) and (3.33) for radiative decays. There
are nonanalytic $m_q^\half$ and $m_q\ln m_q$ chiral corrections which are
responsible for SU(3) violation in chiral preturbation theory. In paper II,
the relevant couplings are obtained from the quark model and they are related
to the constituent quark masses. Hence, SU(3) violation is already incorporated
there. In this section, a comparison between the two different
approaches for SU(3) violation is made.

   We begin with the $D^*\ri D\pi$ decay. When chiral loop corrections are
included, we  recall from Sec. 2 that its decay amplitude is given by
\foot{Chiral logarithmic corrections to the decay $D^{*+}\ri D^0\pi^+$ are
also calculated in Ref.[19] with the up and down quark masses being neglected.
We agree with the results of Ref.[19] if the same approximation is made.}
$$A(D^*_i\ri D_j\pi^a)= \,{g_{\rm eff}\over f_0}\sqrt{ M_{_D}M_{_{D^*}} }(
\lambda^a)_{ij}(\ep^*\cdot q),\eqno(4.1)$$
where $g_{\rm eff}$ has been defined in Eq.(2.30) and $q$ is the pion
momentum. The decay widths implied by the amplitude (4.1) are
$$\eqalign{ \Gamma(D^*\ri D\pi^+) =&\,{1\over 12\pi}\left({g_{\rm eff}\over
f_0}\right)^2{M_{_D}M_{_{D^*}}\over (M^\phys_{_{D^*}})^2}\,q^3,   \cr
\Gamma(D^*\ri D\pi^0) =&\,{1\over 24\pi}\left({g_{\rm eff}\over f_0}\right)^2{
M_{_D}M_{_{D^*}}\over (M^\phys_{_{D^*}})^2}\,q^3.   \cr}\eqno(4.2)$$
Note that $D_s^{*+}\ri D_s^+\pi^0$ is prohibited
by isospin symmetry and $D^{*0}\ri D^+\pi^-$ is kinematically forbidden.
The unrenormalized masses $M_P$ and $M_{P^*}$ appearing in
(2.1), (2.2), (4.1) and (4.2) can be
inferred from the experimental measurement of heavy meson mass differences.
First, the SU(3)-invariant $\alpha_3$ term in Eq.(2.18) can be absorbed into
a redefinition of $M_P$ and $M_{P^*}$. Then, (2.16) and (2.18) yield
$$M_{_{D^+}}^{\phys}-M_{_D} = \,M_{_{D^{*+}}}^{\phys}-M_{_{D^*}}
 =-\alpha_1m_d-{g^2\over 32\pi}\left(3{m^3_\pi\over f^2_0}+2{m^3_\K\over
f^2_0}+{1\over 3}{m^3_\eta\over f^2_0}\right),  \eqno(4.3a)$$
and
$$M_{_{D^+_s}}^{\phys}-M_{_{D}} = \,M_{_{D_s^{*+}}}^{\phys}-M_{_{D
^{*}}} =-\alpha_1m_s-{g^2\over 32\pi}\left(4{m^3_\K\over
f^2_0}+{4\over 3}{m^3_\eta\over f^2_0}\right).  \eqno(4.3b)$$
Eq.(4.3) implies the mass difference
\foot{In the presence of $1/m_Q$ corrections, the heavy quark symmetry
relations $\alpha_2=-\alpha_1$ and $g=f/2$ are no longer valid and Eq.(4.4)
is modified to
$$\eqalign{  M_{_{D^+_s}}^\phys-M_{_{D^+}}^\phys=&\,-\alpha_1(m_s-m_d)-{f^2
\over 128\pi}\left(-3{m_\pi^3
\over f_0^2}+2{m_{_K}^3\over f_0^2}+{m_\eta^3\over f_0^2}\right),  \cr
M_{_{D_s^{*+}}}^\phys-M_{_{D^{*+}}}^\phys=& \,\alpha_2(m_s-m_d)-{1\over 128\pi}
\left({8\over 3}g^2+{1\over 3}f^2\right)\left(-3{m_\pi^3
\over f_0^2}+2{m_{_K}^3\over f_0^2}+{m_\eta^3\over f_0^2}\right).
\cr}$$}
$$
M_{_{D^+_s}}^\phys-M_{_{D^+}}^\phys=\,M_{_{D_s^{*+}}}^\phys-M_{_{D^{*+}}}^
\phys= \,-\alpha_1(m_s-m_d)-{g^2\over 32\pi}\left(-3{m_\pi^3
\over f_0^2}+2{m_{_K}^3\over f_0^2}+{m_\eta^3\over f_0^2}\right).
\eqno(4.4)$$
Note that formula (4.4) is different from that of Ref.[20] by a factor of 4/3
for the second term on the r.h.s. The parameter $\alpha_1$ can be
determined from the
measured value $M^\phys_{_{D_s^+}}-M^\phys_{_{D^+}}
=99.5\pm 0.6$ MeV [21], the current quark masses $m_s=199$ MeV, $m_d=9.9$ MeV
[22], and the value of $g$ chosen. The unrenormalized masses $M_{_D}$ and
$M_{_{D^*}}$ can then be solved from (4.3).
\foot{At the preferred value $g=0.52$ as shown below, we obtain $M_{_D}=1971$
MeV and $M_{_{D^*}}=2112$ MeV.}

    We digress here to comment on the quantity
$$\Delta_{_D}\equiv (M_{_{D_s^{*+}}}^\phys-M_{_{D^{*+}}}^\phys)-(M_{_{D_s^+}}^
\phys-M_{_{D^+}}^\phys)=\,(M_{_{D_s^{*+}}}^\phys-M_{_{D_s^+}}^\phys)-(M_{_{D^
{*+}}}^\phys-M_{_{D^+}}^\phys)\eqno(4.5)$$
with the experimental value [21,23]
$$\Delta_{_D}=\,(0.9\pm 1.9)\,{\rm MeV},\eqno(4.6)$$
which has recently received a lot of attention. Obviously, $\Delta_{_D}$
vanishes in either the heavy quark or chiral limit. Beyond these limits,
$\Delta_{_D}$ receives two types of contributions. The first type is from the
tree contribution induced by the $1/m_c$ corrections to the relation $\alpha_1
=-\alpha_2$. The second type comes from the self-energy one-loop diagrams
with an insertion of lowest order mass splitting $\Delta M^2P(v)
P^\dagger (v)$ and the ones without the above insertion but taking
into account the splitting of $g$ and $f/2$.
This was discussed in detail in our earlier paper [13] on heavy
quark symmetry breaking effects.

   For the radiative decays $D^*\to D\gamma$, we see from
Eq.(2.55) that the leading chiral corrections are dominated by Fig. 2e,
and the wave function and vertex renormalizations.
Before making concrete estimates on the parameter $\rho_i$ [cf. Eq.(2.41)],
recall that in paper II we have considered the constituent quark model and
incorporated  SU(3) breaking into the light quark charge matrix
$$\Q\to \tilde{\Q}=\left(\matrix{ {2\over 3} & 0 & 0  \cr 0 & -{1\over 3}
{M_u\over M_d} & 0 \cr  0 & 0 & -{1\over 3}{M_u\over M_s}  \cr}\right).
\eqno(4.7)$$
To avoid any confusion with the current quark mass $m_q$, we have used the
capital letters to denote the constituent quark mass in (4.7).
Therefore, the ``effective'' parameters $(\beta_{\rm eff})_i$ are identical to
${1\over M_u},~{1\over M_d}$, and ${1\over M_s}$ respectively for
$D_i^*\to D_i+\gamma~(i=1,2,3)$ and hence in the quark model
$$\rho_i^\qmu=\,e\sqrt{M^\phys_{_{D_i}}M^\phys_{_{D^*_i}}}\left({e_i\over M_{
q_i}}+{2\over 3}{1\over m_c}\right).\eqno(4.8)$$
Using the constituent quark masses from the Particle Data Group [21]
$$M_u=\,338\,{\rm MeV},~~M_d=\,322\,{\rm MeV},~~M_s=510\,{\rm MeV},
\eqno(4.9)$$
we get
$$\rho_1^\qmu=\,4.62\,e,~~~\rho_2^\qmu=\,-1.20\,e,~~~\rho_3^\qmu=\,-0.49\,e,
\eqno(4.10)$$
for $m_c=1.6$ GeV. The effective coupling $\rho_i$ enters the radiative
decay rate via
$$\Gamma(D^*_i\to D_i+\gamma)=\,{\rho^2_i\over 12\pi (M_{_{D^*}}^\phys)^2}\,
k^3,\eqno(4.11)$$
with $k$ being the photon momentum in the c.m. frame.

   Recall that
  the quark model predictions agree very well with the two existing data on
the $D^*$ decays [10]: the upper limit on the total width of $D^{*+}$ [12] and
the branching ratio measurements by CLEO II [11]. It would be very important
to see how well chiral perturbation theory does in this regard.
There are four unknown parameters in the theory: $g,~\beta,~m_c$ and
$\Lambda$, and available data do not permit an unambiguous determination of
these parameters. Furthermore, other corrections such as $1/m_Q$ effects may
be of comparable importance as the chiral loop corrections. Unfortunately,
these $1/m_Q$ corrections are difficult to estimate [13]. Consequently, we will
take a modest approach. We will set $m_c=1.6$ GeV and $\Lambda=\Lambda_\chi\sim
1$ GeV, the chiral symmetry breaking scale. Then for $\beta=2.6\,{\rm GeV}^
{-1}$ and $3.0\,{\rm GeV}^{-1}$, we let $g$ take
several different values at $g=0.5,~0.52,~0.60$ and 0.75.
The results are presented in Table I and Table II.
\foot{Strong decay widths are determined from (2.30) and (4.2)
while radiative decay rates are calculated from (2.55) and (4.11).}
We see that for the total width of $D^{*+}$ to be of
order 130 keV or less we must have $g\lsim 0.52$. On the other hand, the
branching ratios for the $D^{*0}$ measured by CLEO II appear to favor
$g=0.52$ and $\beta=2.6\,{\rm GeV}^{-1}$. For this particular choice of $g$
and $\beta$, the predictions are quite close to the quark model results
except for the radiative decay of $D^{*+}_s$. The decay rate for
$D_s^{*+}\ri D_s^+\gamma$ is larger by an order of magnitude in chiral
perturbation theory than in the quark model. Specifically, we find from
Eq.(2.55) that
$$\rho_1=\,4.59\,e,~~~\rho_2=\,-1.14\,e,~~~\rho_3=\,-1.79\,e,\eqno(4.12)$$
which should be compared with the quark model prediction (4.10).
Of course, there is no reason to expect that these two different approaches
for SU(3) violation should agree with each other exactly.
Loosely speaking, SU(3) violation is treated nonperturbatively in the
nonrelativistic quark model, while it is calculated in terms of a
perturbative expansion in chiral effective Lagrangian theory. Nevertheless,
the fact that the predicted decay rates and branching ratios for $D^{*+}$
and $D^{*0}$ agree so well with the quark model and the data suggests a
consistency between ``model and theory". It will be extremely interesting to
measure the rate for $D_s^{*+}\ri D_s^+\gamma$ to see which prediction, if
any, is closer to the truth. It is obvious from Table I that a smaller total
width of the $D^{*+}$ [12] can be easily accommodated without upsetting the
agreement of the predicted branching ratios for both $D^{*+}$ and $D^{*0}$
with the CLEO II data [11].

 At this point it is worthwhile to compare our favored choice of $g=0.52$ and
$\beta=2.6\,{\rm GeV}^{-1}$ in chiral perturbation theory with the
corresponding parameters $g_\qmd$ and $\beta_\qmd$ in the quark model. Recall
that $g_\qmd=0.75$ [3] and $\beta_\qmd=1/M_u=2.96\,{\rm GeV}^{-1}$ [10].
Since $g_\qmd$ is nonperturbative in nature while $g$ is an unrenormalized
parameter in the chiral Lagrangian approach, they are {\it a priori} not the
same. To see their relation, we write down the $D^*\ri D\pi$ amplitude in
the quark model:
$$A(D^*_i\ri D_j\pi^a)=\,{g_\qmd\over f_\pi}\sqrt{M_{_D}^\phys M_{_{D^*}}^
\phys}\,(\lambda^a)_{ij}(\ep^*\cdot q).\eqno(4.13)$$
For this to be in accordance with (4.1) predicted by chiral perturbation
theory, we are led to
$$g_\qmd\approx \,g\,{Z_2\over Z_1}\,{f_\pi\over f_0}\sqrt{ {M_{_D}M_{_{D^*}}
\over M_{_D}^\phys M_{_{D^*}}^\phys} }.\eqno(4.14)$$
Since $Z_2>Z_1,~f_\pi> f_0,~M_{_D}>M_{_D}^\phys$ and $M_{_{D^*}}>M_{_{D^*}}
^\phys$, it is evident that $g<g_\qmd$. By the same token, the relation
between $\beta$ and $\beta_\qmd$ can be understood along the same line.

    We next shift our attention to the heavy baryon sector.
   For the strong decay of heavy baryons, we shall only consider the decay
$\Sigma_c\to\Lambda_c\pi$ which is experimentally seen, although its
rate has not been measured. (It is likely that no other $B_6^{(*)}
\to B_{\bar{3}}\pi$ or $B_6^*\to B_6\pi$ strong decays are kinematically
allowed.) The decay rate for $\Sigma_c\ri\Lambda_c\pi$ is given by [3]
\foot{Note that $f_0$ is replaced by $f_\pi$ in Ref.[3], since there we work
in the tree approximation only.}
$$\Gamma(\Sigma_c^{++}\ri\Lambda_c^+\pi^+)=\,{1\over 16\pi}\,{(g_2)^2_
{\rm eff}\over f_0^2}\,{(M_{_{\Sigma_c}}+M_{_{\Lambda_c}})^2[(M_{_{
\Sigma_c}}-M_{_{\Lambda_c}
})^2-M_\pi^2]^2\over M^2_{_{\Sigma_c}} }\,q,\eqno(4.15)$$
where [see Eq.(3.28)]
$$ (g_2)_{\rm eff}= \,g_2{\sqrt{Z_2(
\Sigma_c)Z_2(\Lambda_c)}\over Z_1'(\Sigma_c\ri\Lambda_c\pi;5a)},\eqno(4.16)$$
which includes chiral loop effects.
  Before examining the SU(3)-breaking effects induced by chiral loops
in the electromagnetic decays of heavy baryons, it is instructive to examine
the tree amplitudes of (3.33) to see what SU(3) violations are expected from
 the constituent quark model.
{}From Eqs.(3.31) and (4.7) we find the following quark-model predictions:
$$\eqalign{ (a_2)^\qmu_{\rm eff}(\Sigma_c^+\ri
\Lambda_c^+\gamma)=&\,a_2^\qmu\left({2\over 3}+{1\over 3}\,{M_u\over M_d}
\right)=\,1.02a_2^\qmu=\,0.62\,e{\rm GeV}^{-1},  \cr
  (a_2)_{\rm eff}^\qmu({\Xi'}_c^+\ri \Xi_c^+\gamma)=&\,a_2^\qmu\left({2\over
3}+{1\over 3}\,{M_u\over M_s}\right)=\,0.88a_2^\qmu=\,0.53\,e{\rm GeV}^{-1},
\cr
(a_2)_{\rm eff}^\qmu({\Xi'}_c^0\ri\Xi_c^0\gamma)= &\,{1\over 3}a^\qmu_2\left(
{M_u\over M_s}-{M_u\over M_d}\right)=\,-0.13a_2^\qmu=\,-0.078\,e{\rm GeV}^{-1
},  \cr}
\eqno(4.17)$$
and
$$\eqalign{  (a_1-8a'_1)_{\rm eff}^\qmu(\Sigma^{*0}_c\ri\Sigma_c^0\gamma)=&\,
a_1^\qmu{M_u\over M_d}-8a'_1=\,1.05a^\qmu_1-8a'_1=-1.45\,e{\rm GeV}^{-1},  \cr
(a_1-8a'_1)_{\rm eff}^\qmu(\Omega_c^*\to\Omega_c
\gamma)=& \,a_1^\qmu{M_u\over M_s}-8a'_1=\,0.66a_1^\qmu-8a'_1=-1.07\,e{\rm
GeV}^{-1},  \cr
(a_1-8a'_1)_{\rm eff}^\qmu({\Xi'}_c^{*0}\ri{\Xi'}_c^0\gamma)= &\,{1\over 2}
a_1^\qmu\left({M_u\over M_d}+{M_u\over M_s}\right)-8a'_1  \cr
 =&\,0.86a_1^\qmu-8a'_1=-1.26\,e{\rm GeV}^{-1},  \cr   (a_1+16a'_1)_{\rm eff}
^\qmu({\Xi'}_c^{*+}\ri{\Xi'}_c^+\gamma)
= &\,3a_1^\qmu\left({2\over 3}-{1\over 3}\,{M_u\over M_s}\right)+16a'_1 \cr
 =&\,1.34a_1^\qmu+16a'_1=-0.49\,e{\rm GeV}^{-1},  \cr
(a_1+16a'_1)_{\rm eff}^\qmu(\Sigma^{*+}_c\ri\Sigma_c^+\gamma)=&\,3a_1^\qmu
\left({2\over 3}-{1\over 3}\,{M_u\over M_d}\right)+16a'_1  \cr
=&\,0.95a_1^\qmu+16a'_1=-0.104\,e{\rm
GeV}^{-1},  \cr}\eqno(4.18)$$
where we have made use of the quark model predictions [10]
$$a_1^\qmu=-{e\over 3}\,\beta_\qmd,~~~a_2^\qmu=\,{e\over 2\sqrt{6}}\,\beta_
\qmd,\eqno(4.19)$$
with $\beta_\qmd=1/M_u$.
The parameter $a'_1$ is predicted by heavy quark symmetry to be [10]
$$a'_1=\,{e\over 12m_c}.\eqno(4.20)$$
{}From the lowest order Lagrangian we have the SU(3) relations of (3.33).
It is thus clear that a measurement of the decay rates of
$$\eqalign{ & (i)~~\Sigma_c^{*0}\to\Sigma_c^0\gamma,~{\Xi'}_c^{*0}\to {\Xi'}
_c^0\gamma,~\Omega_c^*\to\Omega_c\gamma,  \cr & (ii)~{\Xi'}_c^0\to\Xi_c^0
\gamma~{\rm relative~to}~{\Xi'}_c^+\to \Xi_c^+\gamma~{\rm or}~\Sigma_c^+
\to\Lambda_c^+\gamma,   \cr & (iii)~{\Xi'}_c^{*+}\to{\Xi'}_c^+\gamma,~\Sigma
_c^{*+}\to\Sigma_c^+\gamma  \cr}$$
will provide a nice test on SU(3) breaking.

   We are now ready to compare the quark model predictions with those from
chiral perturbation theory using (3.27-3.28) and (3.34) for heavy baryon's
decays. There are six unknown parameters in the theory: $\Lambda,~m_c,~g_1,
{}~g_2,~a_1$ and $a_2$. Unfortunately, there does not exist any experimental
data
for the heavy baryons to guide the choice of these parameters. It is clearly
premature to make an all-out effort to study how the predictions vary with
the parameters. To gain a general idea of how chiral perturbation theory is
compared with the quark model, we minimize the number of parameters by setting
$\Lambda=1$ GeV and $m_c=1.6$ GeV as before, and by making use of the quark
model relations [10]
$$ g_1 =\,{1\over 3}g,~~~g_2=-\sqrt{{2\over 3}}\,g,~~~
a_1= \,-{e\over 3}\beta,~~~a_2=\,{e\over 2\sqrt{6}}\beta. \eqno(4.21)$$
The remaining two parameters $g$ and $\beta$ are finally fixed with the values
favored by the available $D^*$ decay data:
$$g=0.52\,,~~~~\beta=2.6\,{\rm GeV}^{-1}.\eqno(4.22)$$
The relevant wave function and vertex renormalization constants $Z_2,~Z_1'$
and $Z_1''$ for various processes are given by (3.18), (3.25-3.26) and
(3.41-3.47). Numerically,
$$\eqalign{ Z_2(\Lambda_c)=&\,1.40\,,~~~Z_2(\Xi_c)=\,1.83\,,  \cr
Z_2(\Sigma_c)=&\,1.12\,,~~~Z_2(\Xi'_c)=\,1.14\,,~~~Z_2(\Omega_c)=\,1.19\,,
 \cr }\eqno(4.23a)$$
and
$$Z'_1(\Sigma_c\ri\Lambda_c\pi;5a)=\,1.07\,.\eqno(4.23b)$$
{}From (4.15), (4.16) and (4.23) we find that
$$\Gamma(\Sigma_c^{++}\ri\Lambda_c^+\pi^+)=\,1.89\,{\rm MeV}\eqno(4.24)$$
in the chiral Lagrangian approach as compared with 2.45 MeV in the quark model
[3].

Vertex corrections to radiative decays are
$$\eqalign{ Z_1(\Sigma_c^{*++}\ri \Sigma_c^{++}\gamma)=&\,0.77\,,~~Z_1(\Sigma_
c^{*+}\to\Sigma_c^+\gamma)=\,0.62\,,~~Z_1(\Sigma_c^{*0}\to\Sigma_c
^0\gamma)=\,0.92\ ,   \cr
Z_1({\Xi'}_c^{*+}\to{\Xi'}_c^+\gamma)=&\,1.00\,,~~
Z_1({\Xi'}_c^{*0}\to{\Xi'}_c^0\gamma)=\,0.78\,,
{}~Z_1(\Omega_c^*\to\Omega_c\gamma)=\,0.63\,,  \cr
Z_1''(\Sigma_c^{*+}\to\Lambda_c^+\gamma)=&\,0.88\,,~~~~Z_1''({\Xi'}_c^{+*}\to
\Xi_c^+\gamma)=\,0.87\,,   \cr}\eqno(4.25)$$
and
$$\eqalign{ &\delta a_1(\Sigma^{+1(*)}_Q\ri\Sigma_Q^{+1}\,\gamma;6c) =\,
\delta a_1({\Xi'}_Q^{-\half(*)}\ri{\Xi'}_Q^{-\half}\,\gamma;6c)=\,0.24\,e{
\rm GeV}^{-1},\cr  &\delta a_1(\Omega_Q^{(*)}\ri\Omega_Q\,\gamma;6c)=
 \delta a_1(\Sigma^{0(*)}_Q\ri\Sigma_Q^{0}\,\gamma;6c) =\,0.38\,e{\rm GeV}^{
-1},\cr  &\delta a_1(\Sigma_Q^{-1(*)}\ri\Sigma_Q^{-1}\,\gamma;6c) =\,
\delta a_1({\Xi'}_Q^{\half(*)}\ri{\Xi'}_Q^{\half}\,\gamma;6c) =\,
0.11\,e{\rm GeV}^{-1}, \cr &\delta a_2(\Sigma_Q^{0(*)}\ri\Lambda_Q\,\gamma;7c)
=\,-0.08\,e{\rm GeV}^{-1}, \cr      &\delta a_2({\Xi'}_Q^{\half(*)}\ri{\Xi}_
Q^\half\,\gamma;7c) =\,-0.16\,e{\rm GeV}
^{-1},\cr  &\delta a_2({\Xi'}_Q^{-\half(*)}\ri\Xi_Q^{-\half}\,\gamma;7a+7b+7c)
=\,0.15a_2-0.03\,e{\rm GeV}^{-1}. \cr}\eqno(4.26)$$
We get from Eqs.(3.34), (4.23a) and (4.25-4.26) that
$$\eqalign{
(a_1-8a_1')_{\rm eff}(\Sigma_c^{*0}\to\Sigma_c^0\gamma)
=&\,1.08a_1-8a'_1=-1.36\,e{\rm GeV}^{-1},   \cr
(a_1-8a_1')_{\rm eff}(\Omega_c^*\to\Omega_c\gamma)
=&\,1.46a_1-8a'_1=-1.68\,e{\rm GeV}^{-1},  \cr
(a_1-8a_1')_{\rm eff}({\Xi'}_c^{*0}\to{\Xi'}^0_c\gamma)
=&\,1.19a_1-8a'_1=-1.45\,e{\rm GeV}^{-1},   \cr}\eqno(4.27a)$$
and
$$\eqalign{
(a_1+16a_1')_{\rm eff}({\Xi'}_c^{*+}\to{\Xi'}_c^+\gamma)=&
\,1.02a_1+16a'_1=-0.05\,e{\rm GeV}^{-1},   \cr
(a_1+16a_1')_{\rm eff}(\Sigma_c^{*+}\to\Sigma_c^+\gamma)=&
\,1.35a_1+16a'_1=-0.35\,e{\rm GeV}^{-1}.   \cr}\eqno(4.27b)$$
A comparison between (4.27) and (4.18) indicates that the two approaches give
similar results for some processes such as $\Sigma_c^{*0}\ri\Sigma_c^0\gamma$
and ${\Xi'}^{*0}_c\ri {\Xi'}^0_c\gamma$, while they give dramatically
different results for others such as $\Sigma_c^{*+}\ri\Sigma_c^+\gamma$
and ${\Xi'}^{*+}_c\ri {\Xi'}^+_c\gamma$.
For the decay amplitude
$$A(B^*_6\to B_6+\gamma)=\,\eta_2\bar{u}(k_\mu\ep\!\!/-\ep_\mu
k\!\!\!/)\gamma_5u^\mu,\eqno(4.28)$$
with $\eta_2$ being given by Eq.(3.33), the corresponding decay width is [10]
$$\Gamma(B^*_6\to B_6+\gamma)=\,{k\over 48\pi}\,\eta^2_2\left(1-{m^2_f\over
m^2_i}\right)^2(3m_i^2+m_f^2),\eqno(4.29)$$
where $m_i~(m_f)$ is the mass of the initial (final) baryon in the decay.
This formula is presented here for completeness, and
we will not give any predictions for the decay rates of
spin-${3\over 2}$ heavy baryons as their masses are still unknown.

   As for the electromagnetic decays $B_6\to B_{\bar{3}}+\gamma$, Eqs.(3.34),
(3.44) and (4.23-4.26) lead to
$$\eqalign{(a_2)_{\rm eff}(\Sigma_c^+\to\Lambda_c^+\gamma)= &\,1.27a_2=\,0.67
\,e{\rm GeV}^{-1},  \cr
(a_2)_{\rm eff}({\Xi'}_c^+\to\Xi_c^+\gamma)= &\,1.36a_2=\,0.72\,e{\rm GeV}^
{-1}, \cr
(a_2)_{\rm eff}({\Xi'}^0_c\to\Xi_c^0\gamma)= &\,0.10a_2=\,0.05\,e{\rm GeV}^
{-1}.  \cr}\eqno(4.30)$$
Note that ${\Xi'}^0_c\ri\Xi^0_c\gamma$ receives two contributions: one from
Fig. 7a and the other from Fig. 7c and there is a cancellation between the
two. The sign of this amplitude in (4.30) is actually opposite to that in the
quark model [cf. Eq.(4.17)]. With the formula
$$\Gamma(B_6\to B_{\bar{3}}+\gamma)=\,{1\over \pi}\,\eta_1^2k^3,\eqno(4.31)$$
for the radiative decay amplitude
$$A(B_6\to B_{\bar{3}}+\gamma)=\,\eta_1\bar{u}_{\bar{3}}\sigma_{\mu\nu}k^\mu
\epsilon^\nu u_6,\eqno(4.32)$$
the decay rates of various decays are then predicted to be
\foot{All the charmed baryon masses are taken from the Particle Data Group
[21]. For the mass of ${\Xi'}_c$, we employ the hyperfine mass splitting
$m_{_{\Xi'_c}}-m_{_{\Xi_c}}\simeq 100$ MeV derived in Ref.[25].}
$$\eqalign{
\Gamma(\Sigma_c^+\to\Lambda_c^+\gamma)= &\,112\,{\rm keV},  \cr
\Gamma({\Xi'}_c^+\to\Xi_c^+\gamma)= &\,29\,{\rm keV},  \cr
\Gamma({\Xi'}_c^0\to\Xi_c^0\gamma)= &\,0.15\,{\rm keV},  \cr}\eqno(4.33)$$
which should be compared with the corresponding quark-model results: 93 keV,
16 keV and 0.3 keV.

   The predicted SU(3) breaking patterns in chiral perturbation theory for
radiative decays of charmed mesons and baryons are [see (4.12) and (4.27)]:
$$\eqalign{  &\sum_{\rm pol}\left|A(D^{*+}_s\ri D^+_s\gamma)\right|^2 >
\sum_{\rm pol}\left|A(D^{*+}\ri D^+\gamma)\right|^2,  \cr
&\sum_{\rm pol}\left|A(\Omega_c^*\ri \Omega_c\gamma)\right|^2 > \sum_{\rm pol}
\left|A(\Sigma_c^{*0}\ri \Sigma_c^0\gamma)\right|^2 \sim   \sum_{\rm pol}\left
|A({\Xi'}_c^{*0}\ri {\Xi'}_c^0\gamma)\right|^2,   \cr
& \sum_{\rm pol}\left|A(\Sigma_c^{*+}\ri \Sigma_c^+\gamma)\right|^2 >> \sum_{
\rm pol}\left|A({\Xi'}_c^{*+}\ri {\Xi'}_c^+\gamma)\right|^2, \cr}\eqno(4.34)$$
which are opposite to what expected from the quark model [see (4.10) and
(4.17-4.18)]. It is thus extremely important to measure the decay rates of
the above-mentioned modes to test the underlying mechanism of SU(3) violation.

    We should reiterate the tentative nature of the discussion in this
section. In the heavy meson case, the data for $D^*$ are useful but not
decisive in selecting the values of the parameters. Other important
effects such as $1/m_Q$ corrections have not been incorporated.
The procedure employed here is by no means a best fit. The apparent agreement
between the quark model and chiral perturbation theory is pleasing, but it
should be taken with caution. In the heavy baryon sector, the situation is even
more difficult. First of all, we do not have any experimental data. Secondly,
there are many more parameters here than in the heavy meson sector. The choice
of parameters based on (4.19)-(4.21) is necessarily {\it ad hoc}. The only
virtue is that it provides a rule to fix the parameters. It will not be
surprising if nature picks a very different choice from the present one.
Thirdly, the renormalization effects are very significant, as evident from
(4.23) and (4.24). We may question the validity of chiral perturbation
expansion.
\foot{The similar problem in the meson sector is less severe except for the
$D^{*+}_s$ decay. For $g=0.52$, we have $Z_2(D^{*0})=Z_2(D^{*+})=1.18$, $Z_2(
D^{*+}_s)=1.32$ and all the $Z_1$'s are very close to unity for the strong
decays, and $Z_1(D^{*0}\ri D^0\gamma)=0.74,~Z_1(D^{*+}\ri D^+\gamma)=0.87$,
$Z_1(D^{*+}_s\ri D^+_s\gamma)=0.57$ for radiative decays.}
Fourthly, large cancellations occur sometimes. For all these reasons,
the results obtained in this section should not be considered final. Rather,
they make it clear that more experimental and theoretical works are needed.

\vskip 0.4cm
\noindent {\bf 5.~~~Conclusions}
\vskip 0.2cm
   As we have emphasized before, the lowest-order low-energy dynamics of heavy
 mesons and
heavy baryons is completely determined by the heavy quark symmetry and chiral
symmetry, supplemented by the quark model. In order to gain a different
perspective of our previous quark model calculations, we have investigated in
the present paper chiral symmetry violation induced by
the light current quark masses. Chiral corrections to the strong and
electromagnetic decays of heavy hadrons are computed at the one loop order.
The leading chiral-loop effects are nonanalytic in the forms of $m_q^\half$
and $m_q\ln m_q$. We have also confirmed in chiral perturbation theory at
the one loop level the exact result in QCD that the heavy quark contributions
to the M1 transitions of heavy hadrons are not modified by light quark
dynamics [see (2.51) and (3.35)]. Moreover, all consequences of heavy quark
symmetry are not affected by chiral corrections.
We then applied our results to the radiative
and strong decays of charmed mesons and charmed baryons.

  We found in the meson sector that for a particular set of the parameters
$g$ and $\beta$ inferred from the measured branching ratios of the $D^{*0}$
and $D^{*+}$ and the upper limit on the $D^{*+}$ rate,
the predictions of the strong and electromagnetic decays of the
$D^*$ mesons in the chiral Lagrangian approach are not very different from the
quark model results except for the radiative decay of the
$D^{*+}_s$. Using the quark model to relate the four unknown parameters
in the baryon sector to $g$ and $\beta$, we have computed the
strong decay $\Sigma_c\ri\Lambda_c\pi$ and the radiative decays $B_6^*\ri B_6
+\gamma$ and $B_6\ri B_{\bar{3}}+\gamma$. We found that the chiral-Lagrangian
and quark-model approaches in general give similar results for many processes,
whereas they yield drastically different predictions for others such as
$\Sigma_c^{*+}\ri\Sigma_c^+\gamma$ and ${\Xi'}_c^{*+}\ri{\Xi'}_c^+\gamma$.
Moreover, the predicted SU(3)-breaking patterns as shown in (4.34) are opposite
to the quark model expectations. It is thus of great importance to measure the
decay modes listed in (4.34) to test the underlying mechanism of SU(3)
violation.

   A more meaningful and improved comparison between theory and experiment
has to wait until more data on heavy hadrons become available. In the
meantime, it is important to have a better theoretical understanding of the
corrections due to the $1/m_Q$ effects. We believe that both SU(3) symmetry
breaking and the $1/m_Q$ effects should be considered simultaneously. We will
return to this effort sometime in the future.

\vskip 3.0 cm
\centerline{\bf Acknowledgments}
\bigskip

    One of us (H.Y.C.) wishes to thank Professor C. N. Yang and the
Institute for Theoretical Physics at Stony Brook for their hospitality
during his stay there for sabbatical leave.
T.M.Y.'s work is supported in part by the
National Science Foundation.  This research is supported in part by the
National Science Council of ROC under Contract Nos.  NSC83-0208-M001-012,
NSC83-0208-M001-014, NSC83-0208-M001-015 and NSC83-0208-M008-009.

\endpage

\centerline{\bf APPENDIX A}

In this Appendix we list some of the useful Feynman integrals which are
relevant to this paper:
$$\int {\Lambda^\epsilon d^nl\over (2\pi)^n}\,{l_\mu\over (l^2-m^2+i\epsilon)
(v\cdot
l+i\epsilon)}= iv_\mu\,{m^2\over 16\pi^2}\ln{\Lambda^2\over m^2}, \eqno(A1)$$
$$\int {\Lambda^\epsilon d^nl\over (2\pi)^n}\,{l_\mu l_\nu\over (l^2-m^2+i
\epsilon)^2(v\cdot
l+i\epsilon)}= -i{m\over 16\pi}(g_{\mu\nu}-v_\mu v_\nu),\eqno(A2)$$
$$\int {\Lambda^\epsilon d^nl\over (2\pi)^n}\,{l_\mu l_\nu\over (l^2-m^2+i
\epsilon)(v\cdot
l+i\epsilon)^2}= -i(g_{\mu\nu}-2v_\mu v_\nu)\,{m^2\over 16\pi^2}\ln{\Lambda^2
\over m^2},\eqno(A3)$$
$$\eqalign{  \int &{\Lambda^\epsilon d^nl\over (2\pi)^n}\,{l_\mu l_\nu\over
(l^2-m^2+i\epsilon)
[v\cdot (l+\k)+\delta m+i\epsilon)]}= -i{m^3\over 24\pi}(g_{\mu\nu}-v_\mu
v_\nu)\cr   & +i{1\over 16\pi^2}\ln{\Lambda^2\over m^2}\bigg\{
(g_{\mu\nu}-2v_\mu v_\nu)(v\cdot \k+\delta m)m^2-2(g_{\mu\nu}-4v_\mu v_\nu)
(v\cdot\k)\delta m^2 \bigg\}  \cr
& +\,i{m\over 16\pi}(g_{\mu\nu}-3v_\mu v_\nu)(v\cdot\k)\delta m,
  \cr}\eqno(A4)$$
$$\eqalign{
\int {\Lambda^\epsilon d^nl\over (2\pi)^n}\,& {l_\mu l_\nu l_\alpha\over
[(l+k)^2-m^2+i\epsilon]
(l^2-m^2+i\epsilon)(v\cdot l+i\epsilon)}~~({\rm with}~k^2=0~{\rm and}~v\cdot
k=0)  \cr
&=i{m\over 32\pi}\left\{\big[(g_{\mu\nu}-v_\mu v_\nu)k_\alpha+~{\rm
cyclic~in} ~\mu,\nu,\alpha\,\big]+{1\over 2m^2}k_\mu k_\nu k_\alpha\right\}
\cr
&+i{m^2\over 32\pi^2}\ln{\Lambda^2\over m^2}\left\{\big[ (g_{\mu\nu}+{2\over
3m^2}k_\mu k_\nu)v_\alpha+~{\rm cyclic~in}~\mu,\nu,\alpha\,\big]-2v_\mu
v_\nu v_\alpha\right\},   \cr}
 \eqno(A5)$$
where the common factor
$${1\over \epsilon^{\prime}}\equiv {1\over \epsilon}-{1\over 2}\gamma_{_{\rm
E}}+{1\over 2}\ln4\pi,\eqno(A6)$$
with $\epsilon=4-n$ has been lumped into the logarithmic term $\ln{\Lambda^2
\over m^2}$ as $\Lambda$ is an arbitrary  renormalization scale occurred in
the dimensional regularization approach. Note that only the leading
contributions linear in $v\cdot\k$ are retained in (A4).
Our results for the integrals (A3) and (A4) are in agreement with Ref.[17].
Eq. (A5) is derived by assuming $v\cdot k=0$, where $k$ is the
momentum of the outgoing photon which couples to heavy hadrons. This is a
legitimate assumption as long as both incoming and outgoing hadrons are on
shell.
To evaluate above integrals, we first apply tensor decomposition as well as
momentum expansions to reduce them into the following prototypes:
$$I_1(\alpha)=\int {\Lambda^{\epsilon}d^nl\over (2\pi)^n}{1\over (l^2-m^2+i
\epsilon)^{\alpha}(v\cdot l+i
\epsilon)},\eqno(A7)$$
and
$$I_2(\alpha)= \int {\Lambda^{\epsilon}d^nl\over (2\pi)^n}{1\over (l^2-m^2+i
\epsilon)^{\alpha}(v\cdot l+i
\epsilon)^2}.\eqno(A8)$$
We observe that all the external momenta have disappeared from the denominator
due to the expansion:
$${1\over (l+k)^2-m^2}={1\over l^2-m^2}\left(1-{2l\cdot k+k^2\over l^2-m^2}
\right)+\cdots. \eqno(A9)$$
This expansion is a valid procedure since the momentum $k$ is very soft in the
current context. Similar type of expansions can be applied to Eq. (A4) for
bringing up the residual momentum $\k$ to the numerator.

To carry out integrals $I_1$ and $I_2$, the usual Feynman parametrization
$${1\over a^mb^n}=\,{\Gamma(m+n)\over\Gamma(m)\Gamma(n)}\int_0^1{x^{m-1}(1-x)
^{n-1}dx\over [ax+b(1-x)]^{m+n}}\eqno(A10)$$
does not allow one to shift the loop momentum $l$ without destroying the
structure of the integral. Hence, it is convenient to combine denominators
using the identity
$${1\over a^mb^n}=\,{\Gamma(m+n)\over\Gamma(m)\Gamma(n)}\int^\infty_0{2^n
\lambda^{n-1}d\lambda\over (a+2\lambda b)^{m+n}}.\eqno(A11)$$
which is obtained from (A10) by changing variable $x={1\over 1+2\lambda}$.
Sometimes the integration can even be done without combining denominators.
We consider the integral $I_1$ as an example to illustrate the importance
of the $i\epsilon$ terms presented in the denominator. We write
$$\eqalign{I_1(\alpha)&=\int {\Lambda^{\epsilon}
d^nl\over (2\pi)^n}{1\over (l^2-m^2+i\epsilon)^{\alpha}(v\cdot l+i
\epsilon)}\cr
&={1\over 2}\int {\Lambda^{\epsilon}d^nl\over (2\pi)^n}{1\over (l^2-m^2+i
\epsilon)^{\alpha}}
\left({1\over v\cdot l+i\epsilon}+{1\over -v\cdot l+i\epsilon}\right ).\cr}
\eqno(A12)$$
The second line follows from the fact that $I_1$ is invariant under the
substitution $l\to -l$.  Since
$${1\over v\cdot l+i\epsilon}={\cal P}{1\over v\cdot l}-i\pi\delta(v\cdot
l),\eqno(A13)$$
the integral becomes
$$I_1(\alpha)=-i\pi\int{\Lambda^{\epsilon}d^nl\over (2\pi)^n}\delta(v\cdot l){
1\over
(l^2-m^2+i\epsilon)^{\alpha}}.\eqno(A14)$$
Since $I$ is a scalar integral, we can easily evaluate it by going to the rest
frame of $v$.
In the rest frame of $v$, $v^\mu=(1,~\vec{0})$, we find
$$I_1(\alpha)=-i\pi(-1)^{\alpha}\Lambda^{\epsilon}\int {d^{n-1}l\over (2\pi)^n
}{1\over ({\vec l}^2+m^2-i\epsilon)^{\alpha}},\eqno(A15)$$
where $\vec l$ is a vector in the ($n-1$)-dimensional space.
Performing the angular integration gives
$$I_1(\alpha)=-{i\pi(-1)^{\alpha}\Lambda^{\epsilon}\over (2\pi)^n}\cdot {2\pi^
{n-1\over 2}\over \Gamma ({n-1\over 2})}\int dl{l^{n-2}\over (l^2+m^2)^
{\alpha}}.\eqno(A16)$$
Now the $l$-integration is standard, which gives
$$I_1(\alpha)=-{i\pi(-1)^{\alpha}\Lambda^{\epsilon}\over (2\pi)^n}\cdot \pi^{
n-1\over 2}{\Gamma (\alpha-{n-1\over 2})\over \Gamma (\alpha)}(m^2)^{{n-1\over
 2}-\alpha}.\eqno(A17)$$
It is easy to see that $I_1(\alpha)$ is finite for any integer $\alpha$.
Taking $n=4$, we have
$$I_1(\alpha)=-{i(-1)^{\alpha}\sqrt{\pi}\over 16\pi^2}\,{\Gamma (\alpha-{3
\over 2})\over \Gamma (\alpha)}(m^2)^{{3\over 2}-\alpha}.\eqno(A18)$$
To evaluate $I_2(\alpha)$, it is more convenient to apply the identity (A11)
since
the principal part of the integral no longer cancels.
By the power counting argument, one expects $I_2(1)$ to be logarithmically
divergent. However, as we will show momentarily,
such ultraviolet divergence will appear in the $\lambda$ integration rather
than in the momentum one. Therefore some care is needed in order to
consistently treat the infinities. To illustrate this, let us first apply Eq.
(A11) to combine denominators of $I_2(1)$. This procedure
gives:
$$I_2(1)=\int_0^{\infty} 8\lambda d\lambda\int \Lambda^{\epsilon}{d^nl\over
(2\pi)^n} {1\over [l^2-(m^2+\lambda^2)]^3}.\eqno(A19)$$
It is obvious that no infinities arise in the momentum integration. Carrying
out the momentum integration, we obtain
$$I_2(1)=-2i\pi^2\Lambda^{\epsilon}\int_0^{\infty} d\kappa{1\over \kappa +m^2},
\eqno(A20)$$
where we have changed the variable such that $\kappa=\lambda^2$. Apparently
the ultraviolet divergence now appears in the $\kappa$ integration.
To consistently implement the dimensional regularization, let us perform an
analytic continuation on $I_2(1)$ with the aid of
$$\int d^nl {\partial\over \partial l_{\mu}}(l_{\mu} f(l))=\int d^nl \cdot
l_{\mu} {\partial\over \partial l_{\mu}}f(l)+n\int d^n l f(l),\eqno(A21)$$
where $f$ is an arbitrary function of $l$.
The l.h.s. of Eq. (A21) can be set to zero since the integrand is a total
derivative. Now $I_2(1)$ can be defined in terms of $I_2(2)$ through
Eq. (A21):
$$I_2(1)=-{2m^2\over 4-n}I_2(2).\eqno(A22)$$
Since $I_2(2)$ is convergent, the infinity in $I_2(1)$  manifests itself as a
simple pole at $n=4$.
To determine $I_2(1)$ including its finite parts, one has to evaluate
$I_2(2)$ to ${\cal O}(4-n)$. Applying Eq. (A11) gives
$$I_2(2)=\int_0^{\infty}24\lambda d\lambda \int {\Lambda^{\epsilon}d^n l\over
 (2\pi)^n} {1\over [l^2-(m^2+\lambda^2)]^4}.\eqno(A23)$$
Carrying out both integrations, we arrive at
$$I_2(2)={2i\over 16\pi^2m^2}\left[1-{\epsilon\over 2}(\gamma_{_{\rm E}}-\ln
\pi-2\ln2)+{\epsilon\over 2}\ln{\Lambda^2\over m^2}\right],\eqno(A24)$$
By Eq. (A21), we obtain
$$I_2(1)=-{4i\over 16\pi^2}\left[{1\over \epsilon}-{1\over 2}(\gamma_{_{\rm
E}}-\ln\pi-2\ln2)+{1\over 2}\ln{\Lambda^2\over m^2}\right].\eqno(A25)$$
To compute any $I_2(\alpha)$ with $\alpha>2$, one may repeatedly apply Eq.
(A14). The general result is given by
$$I_2(\alpha)={i\over 16\pi^2m^2}{2(-m^2)^{2-\alpha}\over \alpha-1},
\eqno(A26)$$
with $\alpha\geq 2$.

 Finally, the integral (A5) can also be carried out by combining the first two
denominators via the conventional Feynman parametrization (A10), followed by
another combination with the third denominator $(v\cdot l+i\epsilon)$ using
the identity (A11). As a consequence,
$$\eqalign{ &{1\over [(l+k)^2-m^2+i\epsilon] (l^2-m^2+i\epsilon)(v\cdot l+i
\epsilon)} =\,4\int^\infty_0d\lambda\int^1_0 dx  \cr
&\times {1\over[(l+v\lambda+kx)^2+x(1-x)k^2
-\lambda^2-m^2]^3}.  \cr}\eqno(A27)$$
After the momentum integration, the integration over $x$ becomes trivial.
\vskip 1.5 cm
\centerline{ {\bf APPENDIX B}}
\def\ket#1{|#1\rangle}
\def\bra#1{\langle#1|}
In this Appendix we explain the procedures used in Section 3 to obtain the
results for strong and electromagnetic vertices, especially how the SU(3)
group factors are defined and arrived at.  We will consider two examples:
one for a strong vertex and one for an electromagnetic vertex.  For the
strong vertex, we will take Fig.~4a with sextet baryons as the intermediate
states.  The corresponding tree amplitude can be written as
$$
A (S_{ij}^\nu \rightarrow S_{kl}^\mu + \pi^a) = i {3\over 4}\, {g_1\over f_0}
\epsilon_{\mu \alpha \beta \nu}\, \ov{\cal{U}}^\mu v^\alpha q^\beta
{\cal{U}}^\nu \bra{S_{kl}}\tr (B^{\dag}\lambda^a B)~\ket{S_{ij}},
\eqno(B1)
$$
where $B$ is the sextet baryon matrix as defined in Eq.(3.2);
$\ket{S_{ij}}$ and $\ket{S_{kl}}$ are respectively
the SU(3) flavor wave functions for
the initial and final baryons, such that
$$\eqalign{
&B_{mn}\ket{S_{ij}}={1\over\sqrt{2}}
(\delta_{mi}\delta_{nj}+\delta_{mj}\delta_{ni})\ket{0}~~~(i\neq j),\cr
&B_{mn}\ket{S_{ii}}=\delta_{mi}\delta_{ni}\ket{0}.\cr}
\eqno(B2)
$$
Hence for the specific case of
$S_{ij} \rightarrow S_{ij} + \pi^3$, we have
$$
A(S^\nu_{ij} \rightarrow S^\mu_{ij} + \pi^3) = i {3\over 8}\,
{g_1\over f_0} \epsilon_{\mu \alpha \beta \nu}\, \ov{\cal{U}}^\mu v^\alpha
q^\beta {\cal{U}}^\nu (\lambda^3_{ii} + \lambda^3_{jj})~.
\eqno(B3)
$$
The loop contribution from Fig. 4a with sextet intermediate states
can be written in the following form
$$
A_S (S_{ij}^\nu \rightarrow S_{kl}^\mu + \pi^a)
= \sum_b \ov{\cal{U}}^\mu M_{\mu\nu}
(b) ~{\cal{U}}^\nu \bra{S_{kl}}G (b)\ket{S_{ij}}~,
\eqno(B4)
$$
where $M_{\mu\nu} (b)$ is the contribution from the loop
excluding the SU(3) matrices:
$$
M_{\mu \nu} (b)  =  \left( -i {3g_1\over 4f_0} \right)^3 \int {d^4
\ell\over (2\pi)^4} \epsilon_{\mu \sigma \lambda \kappa} \ell^\lambda v^\kappa
\epsilon_{\alpha \gamma \beta \delta} q^\beta v^\delta
\epsilon_{\rho \nu \xi \zeta} (-\ell^\xi) v^\zeta
$$
$$
\times{i(-g^{\sigma \alpha} + v^\sigma v^\alpha)\over v\cdot \ell} \cdot{i (-g
^{\rho\gamma} + v^\rho v^\gamma)\over v \cdot \ell}\cdot{i\over \ell^2 -m^2_{
\pi^b}}~~,\eqno(B5)
$$
where we have neglected the residual momentum $\tilde{k}$ of the
initial baryon and the external pion momentum $q$ in the propagators since
we are only interested in the leading order contribution.  Using the result
(A3) for the integral, we obtain
$$
M_{\mu \nu} (b) = -2i \epsilon_{\mu \alpha \beta \nu} v^\alpha q^\beta
{1\over f_0} \left( {3\over 4} g_1 \right)^3 \epsilon_{\pi^b}~,
\eqno(B6)
$$
where $\epsilon_{\pi^b}$ is defined in (2.22).  The operator
$G(b)$ is a product of traces involving
SU(3) matrices and the sextet baryon matrix $B$:
$$
G(b)=\tr(B^\dagger\lambda^b B)\cdot \tr(B^\dagger\lambda^a B)\cdot
     \tr(B^\dagger\lambda^b B).
\eqno(B7)
$$
Note that the adjacent intermediate $B$'s and $B^{\dag}$'s in (B7)
combine in pairs to form
sextet ``propagators" in the SU(3) flavor space, namely
$$
\VEV{B_{k\ell} {B^\dagger}_{mn}}
= {1\over 2} \left( \delta_{kn} \delta_{\ell m}
+ \delta_{km} \delta_{\ell n} \right)~.
\eqno(B8)
$$
Thus $G(b)$ can be rewritten as
$$
G(b) = {1\over 4} \tr \left\{ {B^\dagger} \lambda^b \lambda^b B \lambda^{aT} +
{B^\dagger} \lambda^b \lambda^a B \lambda^{bT}  + {B^\dagger} \lambda^b B
(\lambda^a \lambda^b)^T + {B^\dagger} \lambda^b \lambda^a \lambda^b B \right\},
\eqno(B9)
$$
where the superscript $T$ signifies the transposition of a matrix.
Again, for the case of $S_{ij} \rightarrow S_{ij} + \pi^3$,
the flavor matrix element can be easily worked out to be
$$
\bra{S_{ij}} G(b) \ket{S_{ij}} = \xi_{ij}~~,
\eqno(B10)
$$
where $\xi_{ij}$ is given by (3.20).
When (B3) and (B10) are combined, we obtain
$$
A_S (S^\nu_{ij} \rightarrow S^\mu_{ij} + \pi^3 ) = -2i {1\over f_0}
\epsilon_{\mu \alpha \beta \nu}\, \ov{\cal{U}}^\mu v^\alpha q^\beta
{\cal{U}}^\nu\left({3\over 4} g_1\right)^3 \sum_b \epsilon_{\pi^b} \xi_{ij}~~.
\eqno(B11)
$$
Adding up the tree amplitude (B3), we find
$$
A(S^\nu_{ij} \rightarrow S^\mu_{ij} + \pi^3) = i {eg_1\over 8 f_0}\,
\epsilon_{\mu \alpha \beta \nu}\, \ov{\cal{U}}^\mu v^\alpha q^\beta
{\cal{U}}^\nu (\lambda^3_{ii} + \lambda^3_{jj})
\left[ 1 - {9\over 4} g^2_1 \sum_b {\epsilon_{\pi^b}
\xi_{ij}\over \lambda^3_{ii} + \lambda^{3}_{jj}} \right]~~.
\eqno(B12)
$$
The quantity in the square brackets of (B12) is the contribution
to $Z^{-1}_1$ due to the intermediate sextet baryons of Fig.~4a.  It agrees
with the term proportional to $g^2_1$ of Eq.(3.19).

The contribution to $Z_1$ due to antitriplet baryon intermediate states of
Fig.~4a can be computed similarly.  The SU(3) flavor ``propagator'' for an
antitriplet is
$$
\VEV{T_{k \ell} {T^\dagger}_{mn} } = \delta_{kn} \delta_{\ell m} - \delta_{km}
\delta_{\ell n}~~.
\eqno(B13)
$$
The difference in normalization between (B13) and (B8) is a
result of how the matrices for a sextet and an antitriplet are defined [see
(3.1) and (3.2)~].  The minus sign in (B13) accounts for the
antisymmetry of the antitriplet baryon matrix $T$ (or $B_{\bar 3}$).

We now give an example of the calculation for a loop diagram in which a
photon couples to Goldstone bosons.  Consider the Feynman diagram of
Fig.~6c with intermediate sextet baryons.  The interactions of the
Goldstone bosons with the electromagnetic vector potential $A_\mu$ are
introduced into Eq.(2.20) by the substitution
$$
\partial_\mu \Sigma \rightarrow D_\mu \Sigma \equiv \partial_\mu \Sigma + ie
A_\mu [{\cal{Q}}~,~\Sigma ]~~.
\eqno(B14)
$$
The electromagnetic vertex for $\pi^a \rightarrow \pi^b + \gamma$
is given by
$$
A(\pi^a \rightarrow \pi^b + \gamma) = -ie (2q-k) \cdot \varepsilon
\left[- {1\over 2} \tr (\lambda^a [{\cal{Q}}~,~\lambda^b]) \right]~,
\eqno(B15)
$$
where $q$ and $k$ are the momentum of $\pi^a$ and the photon,
respectively.  The quantity in the square brackets is equal to one when the
Goldstone boson has a unit positive charge.  It will be part of the group
factor defined below.  Fig.~6c with intermediate
sextet baryons gives a contribution to $S \rightarrow S + \gamma$:
$$
A_S (S_{ij}^\nu \rightarrow S_{kl}^\mu + \gamma) = \sum_b \ov{\cal{U}}^\mu
M^\prime_{\mu \nu} (b) ~{\cal{U}}^\nu \bra{S_{kl}}G^\prime (b)\ket{S_{ij}},
\eqno(B16)
$$
in a notation similar to (B4).  We find
$$
M^\prime_{\mu \nu} (b) = \left(-i {3 g_1\over 4 f_0} \right)^2 (-ie)
\int {d^4 \ell\over (2 \pi)^4} \epsilon_{\mu \beta \lambda \kappa} (\ell +
k)^\lambda v^\kappa \times
$$
$$
(-2 \ell - k) \cdot \varepsilon\, \epsilon_{\alpha \nu \gamma \delta} (-
\ell^\gamma) v^\delta {i\over (\ell + k)^2 -m^2_{\pi^b}} \cdot
{i\over \ell^2 -m^2_{\pi^b}}\cdot {i (-g^{\alpha \beta} + v^\alpha
v^\beta)\over v \cdot \ell}~~.
\eqno(B17)
$$
Notice that the Goldstone bosons $\pi^a$ and $\pi^b$ on both
sides of the electromagnetic vertex must have the same mass which we denote
as $m_{\pi^b}$.  The integrals needed in (B17) are given in Appendix~A.
After neglecting a contribution to the convection current proportional to
$v \cdot \varepsilon$, we get the gauge invariant result
$$
M^\prime_{\mu \nu} (b) = e \left( {3g_1\over 4f_0} \right)^2
{m_{\pi^b}\over 16 \pi}\, i (k_\mu \varepsilon_\nu - k_\nu \varepsilon_\mu)~~.
\eqno(B18)
$$
The group factor operator $G^\prime (b)$ is given by
$$
G^\prime (b) = \sum_a \tr ({B^\dagger} \lambda^b B)
                \cdot \tr ({B^\dagger} \lambda^a B)
\left[-{1\over 2} \tr \left(\lambda^a [{\cal{Q}}~,~\lambda^b ] \right)
\right]~~.
\eqno(B19)
$$
Making use of the sextet propagator (B8) and the identity
$$
\sum_a \lambda^a_{jk} \lambda^a_{mn} = 2 (\delta_{jn} \delta_{km} -
{1\over 3} \delta_{jk} \delta_{mn})~~,
\eqno(B20)
$$
we obtain
$$
G^\prime (b) = - {1\over 2} \tr \left\{ {B^\dagger} \lambda^b
[{\cal{Q}}~,~\lambda^b] B + {B^\dagger} \lambda^b B [{\cal{Q}}~,~\lambda^b]^T
\right\}~~.
\eqno(B21)
$$
For the case $S_{ij} \rightarrow S_{ij} + \gamma$, we find
$$\eqalign{
\bra{S_{ij}}G^\prime (b) \ket{S_{ij}}
= &- {1\over 4} \Big\{ \left( \lambda^b [{\cal{Q}}~,~\lambda^b]\right)_{ii}
+ \lambda^b_{ii} [{\cal{Q}}~,~ \lambda^b]_{jj} (1-\delta_{ij})  \cr
+ &\,\lambda^b_{ij} [{\cal{Q}}~,~ \lambda^b]_{ji} + (i \leftrightarrow j)
\Big\}.   \cr}\eqno(B22)$$
As a consequence of the fact that ${\cal Q}$ is diagonal, we have
$$
[{\cal{Q}}~,~ \lambda^b]_{jj} = 0~~,
\eqno(B23a)
$$
$$
\lambda^b_{ij} [{\cal{Q}}~,~ \lambda^b]_{ji} + (i \leftrightarrow j) = 0~~.
\eqno(B23b)
$$
Finally,
$$
\bra{S_{ij}}G^\prime (b)\ket{S_{ij}} = \bar{\xi}_{ij}~~,
\eqno(B24)
$$
where $\bar{\xi}_{ij}$ is defined by (3.55).
Combining (B18) and (B24), we
obtain
$$\eqalign{
A_S (S^\mu_{ij} \rightarrow S^\nu_{ij} + \gamma)
= &\,i {3\over 2}\, \ov{\cal{U}}^\nu ({\cal{Q}}_{ii} + {\cal{Q}}_{jj}) (k_\nu
\varepsilon_\mu - k_\mu \varepsilon_\nu) {\cal{U}}^\mu  \cr
\times & \left[ {e\over 32\pi} \sum_b {m_{\pi^b}\over f^2_0} \cdot {3\over 4}
g^2_1 \bar{\xi}_{ij} {1\over {\cal{Q}}_{ii} + {\cal{Q}}_{jj}} \right]~.  \cr}
\eqno(B25)
$$
The quantity in the square bracket gives rise to
the term proportional to $g^2_1$ in (3.54).

\endpage

\centerline{\bf REFERENCES}
\medskip

\item{1.} N. Isgur and M.B. Wise, {\sl Phys. Lett.} {\bf B232}, 113 (1989);
{\sl Phys. Lett.} {\bf B237}, 527 (1990).

\item{2.} M.B. Voloshin and M.A. Shifman, {\sl Yad. Fiz.} {\bf 45}, 463 (1987)
[{\sl Sov. J. Nucl. Phys.} {\bf 45}, 292 (1987)].

\item{3.} T.M. Yan, H.Y. Cheng, C.Y. Cheung, G.L. Lin, Y.C. Lin, and H.L.
Yu, {\sl Phys. Rev.} {\bf D46}, 1148 (1992); see also T.M. Yan, {\sl Chin. J.
Phys.} (Taipei) {\bf 30}, 509 (1992).

\item{4.} M.B. Wise, {\sl Phys. Rev.} {\bf D45}, R2188 (1992).

\item{5.} G. Burdman and J. Donoghue, {\sl Phys. Lett.} {\bf B280}, 287
(1992).

\item{6.} P. Cho, {\sl Phys. Lett.} {\bf B285}, 145 (1992).

\item{7.} P. Cho and H. Georgi, {\sl Phys. Lett.} {\bf B296}, 408 (1992);
{\sl ibid} {\bf B300}, (E)410 (1993).

\item{8.} J.F. Amundson, C.G. Boyd, E. Jenkins, M. Luke, A.V. Manohar, J.L.
Rosner, M.J. Savage and M.B. Wise, {\sl Phys. Lett.} {\bf B296}, 415 (1992).

\item{9.} H.Y. Cheng, C.Y. Cheung, G.L. Lin, Y.C. Lin, T.M. Yan, and H.L. Yu,
{\sl Phys. Rev.} {\bf D46}, 5060 (1992).

\item{10.} H.Y. Cheng, C.Y. Cheung, G.L. Lin, Y.C. Lin, T.M. Yan, and H.L. Yu,
{\sl Phys. Rev.} {\bf D47}, 1030 (1993).

\item{11.} CLEO Collaboration, F. Butler {\it et al}., {\sl Phys. Rev. Lett.}
{\bf 69}, 2041 (1992).

\item{12.} ACCMOR Collaboration, S. Barlarg {\it et al}., {\sl Phys. Lett.}
{\bf B278}, 480 (1992).

\item{13.} H.Y. Cheng, C.Y. Cheung, G.L. Lin, Y.C. Lin, T.M. Yan, and H.L. Yu,
CLNS 93/1192, to appear in Phys. Rev. D (1994).

\item{14.} A. Manohar and H. Georgi, {\sl Nucl. Phys.} {\bf B234}, 189 (1984).

\item{15.} S. Weinberg, {\sl Physica} {\bf 96A}, 327 (1979).

\item{16.} J. Gasser and H. Leutwyler, {\sl Nucl. Phys.} {\bf B250}, 465
(1985).

\item{17.} P. Cho, {\sl Nucl. Phys.} {\bf B396}, 183 (1993).

\item{18.} B. Grinstein, E. Jenkins, A. Manohar, M. Savage, and
M.B. Wise, {\sl Nucl. Phys.} {\bf B380}, 369 (1992).

\item{19.} R. Fleischer, {\sl Phys. Lett.} {\bf B303}, 147 (1993).

\item{20.} J.L. Goity, {\sl Phys. Rev.} {\bf D46}, 3929 (1992).

\item{21.} Particle Data Group, {\sl Phys. Rev.} {\bf D45}, S1 (1992).

\item{22.} C.A. Dominguez and E.de Rafael, {\sl Ann. Phys. (N.Y.)} {\bf 174},
372 (1987); J. Gasser and H. Leutwyler, {\sl Phys. Rep.} {\bf 87}, 77 (1982).

\item{23.} CLEO Collaboration, D. Bortoletto {\it et al.}, {\sl Phys. Rev.
Lett.} {\bf 69}, 2046 (1992).

\item{24.} L. Randall and E. Sather, {\sl Phys. Lett.} {\bf B303}, 345 (1993);
J.L Rosner and M.B. Wise, {\sl Phys. Rev.} {\bf D47}, 343 (1993);
J.L. Goity, {\sl Phys. Lett.} {\bf B303}, 337 (1993); E. Jenkins,
CERN-TH.6765/92 (1992).

\item{25.} M.J. Savage and M. Wise, {\sl Nucl. Phys.} {\bf B326}, 15 (1989).

\endpage

\centerline{\bf Figure Captions}
\vskip 0.6cm
\item{\rm Fig.~1.} Chiral-loop diagrams contributing to the strong decay
$P^*\ri
P\pi^a$. For simplicity, the wave-function renormalization and mass
counterterms are not shown in Figs. 1-8, but
the necessary mass renormalization is to be understood.
\item{\rm Fig.~2.} Chiral-loop diagrams contributing to the radiative decay
$P^*\ri P\gamma$.
\item{\rm Fig.~3.} Amplification of Fig. 2e with all charged meson loops
specified.
\item{\rm Fig.~4.} Chiral-loop diagrams contributing to the strong decay $S\ri
S\pi^a$. In 4(a), the external pion may originate from an $SS\pi$ or
$ST\pi$ vertex.
\item{\rm Fig.~5.} Chiral-loop diagrams contributing to the strong decay $S\ri
T\pi^a$. In 5(a), the external pion may originate from an $SS\pi$ or
$ST\pi$ vertex.
\item{\rm Fig.~6.} Chiral-loop diagrams contributing to the radiative decay
$S\ri S\gamma$. In 6(a), the photon may originate from an $SS\gamma$ or $ST
\gamma$, or $TT\gamma$ vertex with a coupling constant $a_1$ or $a_2$ or
$a'_1$. In 6(c), (d) and (e), the intermediate state can be an $S$ or $T$
baryon. A diagram similar to 6(b) but with a photon attached to the light
meson does not contribute to the M1 transition of $S\ri S\gamma$.
\item{\rm Fig.~7.} Chiral-loop diagrams contributing to the radiative decay
$S\ri T\gamma$. In 7(a), the photon may originate from an $SS\gamma$ or $ST
\gamma$ vertex with a coupling constant $a_1$ or $a_2$. In 7(c), (d) and (e),
the intermediate state is an $S$ baryon.
\item{\rm Fig.~8.} Chiral-loop diagram contributing to the radiative decay
$T\ri T\gamma$. This is the only loop diagram contributing to the M1
transition $T\ri T\gamma$. All other possible diagrams do not contribute; see
the footnote after Eq.(3.40) for discussion.

\endpage
\centerline{\bf Table Captions}
\vskip 0.6cm
\item{\rm Table~I.} Predicted decay rates (in units of keV) and branching
ratios (in parentheses) of strong and electromagnetic
decays of charmed mesons for various values of $g$ with $\beta=2.6\,{\rm GeV}
^{-1}$ and $m_c=1.6$ GeV. For comparison, the quark model predictions [10]
and the experimental branching ratios measured by CLEO II [11] are given in
the last two columns. The numbers under ``quark model" differ somewhat from
those given in Ref.[10] because of the more precise pion masses used here.
\item{\rm Table~II.} Same as Table I except for $\beta=3.0\,{\rm GeV}^{-1}$.

\endpage

\input phyzzx

\hsize= 9 in

\def\ri{\rightarrow}

Table I. Predicted decay rates (in units of keV) and branching
ratios (in parentheses) of strong and electromagnetic
decays of charmed mesons for various values of $g$ with $\beta=2.6\,{\rm GeV}
^{-1}$ and $m_c=1.6$ GeV. For comparison, the quark model predictions [10]
and the experimental branching ratios measured by CLEO II [11] are given in
the last two columns. The numbers under ``quark model" differ somewhat from
those given in Ref.[10] because of the more precise pion masses used here.

$$\vbox{\tabskip=0pt \offinterlineskip
  \def\tabrule{\noalign{\hrule}}
  \halign to \hsize{
  \strut
  \vrule # & \quad#\hfil~ & \vrule # & ~\quad #\hfil & ~\quad #\hfil
  & \quad #\hfil~  & ~\quad # \hfil & \vrule #
  & \quad # \hfil~ & \quad #\hfil~  & \vrule #\tabskip=0pt\cr
  \tabrule
  & & & & & & & & & &  \cr
  & Reaction &  & $g=0.5$ & $g=0.52$ & $g=0.6$ & $g=0.75$ &
  & quark~model  & CLEO II &  \cr
  & & & & & & & & & &  \cr
  \tabrule
  & & & & & & & & & & \cr
  &$D^{*+}\ri D^0\pi^+$ && 78.8~(68$\%$) & 87.9~(68$\%$) & 133.4~(68$\%$)
  & 276.0~(68$\%$) &&102~(68$\%$)
  & $(68.1\pm 1.0\pm 1.3)\%$ &  \cr
  & & & & & & & & & & \cr
  & $D^{*+}\ri D^+\pi^0$ && 35.7~(31$\%$) & 39.8~(31$\%$) & 60.5~(31$\%$)
  & 125.0~(31$\%$) && 46~(31$\%$)
  & $(30.8\pm 0.4\pm 0.8)\%$ &  \cr
  & & & & & & & & & & \cr
  & $D^{*+}\ri D^+\gamma$ && 1.9~(1.7$\%$) & 2.0~(1.5$\%$) & 2.1~(1.1$\%$)
  & 2.5~(0.6$\%$) && 2~(1.3$\%$) & $(1.1\pm 1.4\pm 1.6)\%$ &  \cr
  & & & & & & & & & & \cr
  & $D^{*+}\ri$~total && 116.4 & 129.7 & 196.0 & 403.5 && 150 & &  \cr
  & & & & & & & & & & \cr
  & & & & & & & & & & \cr
  & $D^{*0}\ri D^0\pi^0$ && 54.1~(61$\%$) & 60.3~(65$\%$) & 91.5~(76$\%$)
  & 189.3~(90$\%$) && 70~(67$\%$) & $(63.6\pm 2.3\pm 3.3)\%$ &  \cr
  & & & & & & & & & & \cr
  & $D^{*0}\ri D^0\gamma$ && 34.0~(39$\%$) & 33.0~(35$\%$) & 28.9~(24$\%$)
  & 19.9~(10$\%$) && 34~(33$\%$) & $(36.4\pm 2.3\pm 3.3)\%$ & \cr
  & & & & & & & & & & \cr
  & $D^{*0}\ri$~total &&  88.1 & 93.3 & 120.4 & 209.2 && 104 &&  \cr
  & & & & & & & & & & \cr
  & & & & & & & & & & \cr
  & $D_s^{*+}\ri D_s^+\gamma$ && 4.6 & 4.5 & 3.7 & 2.1 && 0.3 &&  \cr
  & & & & & & & & & & \cr
\tabrule
  }}$$

\endpage

Table II. Same as Table I except for $\beta=3.0\,{\rm GeV}^{-1}$.

$$\vbox{\tabskip=0pt \offinterlineskip
  \def\tabrule{\noalign{\hrule}}
  \halign to \hsize{
  \strut
  \vrule # & \quad#\hfil~ & \vrule # & ~\quad #\hfil & ~\quad #\hfil
  & \quad #\hfil~  & ~\quad # \hfil & \vrule #
  & \quad # \hfil~ & \quad #\hfil~  & \vrule #\tabskip=0pt\cr
  \tabrule
  & & & & & & & & & &  \cr
  & Reaction &  & $g=0.5$ & $g=0.52$ & $g=0.6$ & $g=0.75$ &
  & quark~model  & CLEO II &  \cr
  & & & & & & & & & &  \cr
  \tabrule
  & & & & & & & & & & \cr
  &$D^{*+}\ri D^0\pi^+$ && 78.8~(67$\%$) & 87.9~(67$\%$) & 133.4~(67$\%$)
  & 276.0~(68$\%$) &&102~(68$\%$)
  & $(68.1\pm 1.0\pm 1.3)\%$ &  \cr
  & & & & & & & & & & \cr
  & $D^{*+}\ri D^+\pi^0$ && 35.7~(30$\%$) & 39.8~(30$\%$) & 60.5~(31$\%$)
  & 125.0~(31$\%$) && 46~(31$\%$)
  & $(30.8\pm 0.4\pm 0.8)\%$ &  \cr
  & & & & & & & & & & \cr
  & $D^{*+}\ri D^+\gamma$ && 3.4~(3$\%$) & 3.4~(3$\%$) & 3.8~(2$\%$)
  & 4.7~(1$\%$) && 2~(1.3$\%$) & $(1.1\pm 1.4\pm 1.6)\%$ &  \cr
  & & & & & & & & & & \cr
  & $D^{*+}\ri$~total && 117.9 & 131.1 & 197.7 & 405.7 && 150 & &  \cr
  & & & & & & & & & & \cr
  & & & & & & & & & & \cr
  & $D^{*0}\ri D^0\pi^0$ && 54.1~(53$\%$) & 60.3~(56$\%$) & 91.5~(67$\%$)
  & 189.3~(85$\%$) && 70~(67$\%$) & $(63.6\pm 2.3\pm 3.3)\%$ &  \cr
  & & & & & & & & & & \cr
  & $D^{*0}\ri D^0\gamma$ && 47.5~(47$\%$) & 46.6~(44$\%$) & 42.6~(33$\%$)
  & 33.3~(15$\%$) && 34~(33$\%$) & $(36.4\pm 2.3\pm 3.3)\%$ & \cr
  & & & & & & & & & & \cr
  & $D^{*0}\ri$~total &&  101.6 & 106.9 & 134.1 & 222.6 && 104 &&  \cr
  & & & & & & & & & & \cr
  & & & & & & & & & & \cr
  & $D_s^{*+}\ri D_s^+\gamma$ && 8.3 & 8.2 & 7.5 & 6.0 && 0.3 &&  \cr
  & & & & & & & & & & \cr
\tabrule
  }}$$

  \end